\documentclass[12pt]{article}

\usepackage[utf8]{inputenc}

\usepackage[top=2cm, bottom=2cm, left=2.3cm, right=2cm]{geometry}
\usepackage{amsmath,amsfonts,amssymb,amsthm}
\usepackage{mathrsfs}  
\usepackage{yhmath}
\usepackage{tikz}
\usepackage{xfp} 

\usepackage[hidelinks]{hyperref}

\usepackage[toc]{appendix}

\usepackage{biblatex} 
\newcommand{\deriv}{\mathrm{d}}
\newcommand{\cte}{C^{te}}

\newcommand{\dvol}{\deriv\mathrm{Vol}^4}

\date{}

\newtheorem{definition}{Definition}[section]
\newtheorem{theorem}{Theorem}[section]
\newtheorem{proposition}{Proposition}[section]

\newtheorem{remark}{Remark}[section]

\begin{document}
\title{Conformal scattering of the wave equation in the Vaidya spacetime.}
\author{Armand Coudray}

\maketitle

\begin{abstract}
We construct the conformal scattering operator for the scalar wave equation on the Vaidya spacetime using vector field methods. The spacetime we consider is Schwarzschild, near both past and future timelike infinities, in order to use existing decay results for the scalar field, ensuring our energy estimates. These estimates guarantee the injectivity of the trace operator and the closure of its range. Finally, we solve a Goursat problem for the scalar waves on null infinities, demonstrating that the range of the trace operator is dense. Consequently, this implies that the scattering operator is an isomorphism.
\end{abstract}

\tableofcontents

\section{Introduction}
The theory of scattering emerges naturally from physics with the objective of characterizing a field by its asymptotic properties, specifically through the observation of distant events (both in terms of distance and time). Constructing a scattering operator is not only a matter of its existence; it is also necessary to prove that the future (or past) behaviour of the field entirely and uniquely characterizes the solution in the rest of the spacetime.
Historically, the study of these asymptotic properties has relied on spectral methods that are ill-suited to generic time dependence. An alternative approach, that authorises the time-dependence of the metric, is to use the concept of conformal compactification, as introduced by Penrose in the 1960s in a series of articles (\cite{1963_Penrose}, \cite{1965_Penrose-Bondi-Hermann}; for a survey of conformal methods, see \cite{1984_Penrose-Rindler}). The physical spacetime $(\mathcal{M},g)$, is embedded into a larger spacetime, denoted by $(\mathcal{\hat{M}},\hat{g})$, which is called the compactified spacetime or un-physical spacetime. $\mathcal{M}$ is the interior of $\hat{\mathcal{M}}$ and the boundary of the compactified spacetime $\partial\hat{\mathcal{M}}$ is made up of two null hypersurfaces, denoted  $\mathscr{I}^{\pm}$, referred to as the future and past null infinities, and three "points", denoted by $i^{\pm}, i^0$, which are the future and past timelike infinities and the spacelike infinity. We obtain $\hat{g}$ using a conformal factor $\Omega$ and putting $\hat{g} = \Omega^2 g$. On $\mathcal{M}$, $\Omega > 0$ and at $\partial\hat{\mathcal{M}},\; \Omega = 0$ and $\nabla_a \Omega \neq 0$. We  also transform the physical field $\phi$, to a rescaled field $\hat{\phi}$, given by $\hat{\phi} = \Omega^{-1}\phi$. For a comprehensive comparison between the two approaches to scattering theory in general relativity, we refer to \cite{2024_Nicolas}.

The first formulation of conformal time-dependent scattering was developed by Friedlander, who observed a direct link between the concept of radiation fields (\cite{1962_Friedlander}, \cite{1964_Friedlander}, \cite{1980_Friedlander}) and the scattering theory of Lax and Phillips (\cite{1967_Lax-Phillips}). The theory of Lax and Phillips uses a translation representer of the solution that corresponds to Friedlander's radiation field, which is an asymptotic profile of the field along outgoing radial null geodesics. The scattering problem is then understood as solving a Goursat problem on $\mathscr{I}$ using the radiation field as data. In 1990, L. Hörmander in \cite{1990_Hormander}, gave a method to solve the Goursat problem for the wave equation on a general spatially compact spacetime, based on energy estimates.

Following this, L. Mason and JP. Nicolas in \cite{2004_Mason-Nicolas} formulated conformal scattering as the construction of a scattering operator on the compactified spacetime that associates the trace of the rescaled field on $\mathscr{I}^+$ to its trace on $\mathscr{I}^{-}$. To ensure the 'good properties' of the scattering operator, it needs to be an isomorphism between past and future null infinities so that the trace of the field on the boundary determines the solution in the interior of the spacetime entirely and uniquely. The geometrical framework of this article was a class of non-stationary vacuum space-times admitting a conformal compactification that is smooth at null and timelike infinities. The extension to black holes was done by JP. Nicolas for the wave equation on the Schwarzschild spacetime in \cite{2016_Nicolas}, the main difficulty being to deal with the singularities at the timelike infinities. The extension possibilities for conformal scattering then took two main directions: either by studying another equation or by modifying the spacetime in which the equation propagates. This was done  on the Kerr geometry in \cite{2004_Hafner-Nicolas}, on Kerr-de Sitter extremal black holes in \cite{2023_Borthwick}, in Reissner-Nordstrom metrics in \cite{2018_Kehle-Shlapentokh-Rothman}, \cite{2021_Hafner-Mokdad-Nicolas}. Note also the extension for non linear equations by Joudioux in \cite{2012_Joudioux} and the application for Maxwell potentials with a particular attention to gauge choices by Taujanskas in \cite{2019_Taujanskas} and by Nicolas and Taujanskas in \cite{2022_Nicolas-Taujanskas}.

In this paper, we construct a scattering theory on a non-empty dynamical spacetime, with a white hole background. The white hole considered is a Schwarzschild white hole with mass $m_{+}$ that evolves to a Schwarzschild white hole with a smaller mass $m_{-}$. The transition from one white hole to the other is described using the Vaidya metric, which models the emission of null dust along null geodesics by the white hole. The Vaidya metric is defined on a given finite retarded time interval to ensure that the past and future timelike infinities are in a Schwarzschild neighbourhood. We need indeed, in order to construct the scattering operator, to prove that the energy of the field is going to zero near $i^{\pm}$. Such decay results are established in the Schwarzschild spacetime by M. Dafermos and I. Rodnianski in \cite{2008_DAfermos_Rodnianski}. As far as the author knows, similar results do not exist on the Vaidya spacetime. This paper could be seen as a part of the analytic study of the wave equation on Vaidya spacetime, other tools can be found in a separate publication by the same author (see \cite{2023_Coudray}).

The paper is organised as follows: Section \ref{section:scattering-geometry} deals with the description of the geometrical framework including definitions and properties of the Schwarzschild and Vaidya spacetimes. In particular we recall the definition of the second optical function  presented in \cite{2021_Coudray-Nicolas} by JP. Nicolas and the author. This function, denoted by $v$ is analogous to the advanced time of Eddington-Finkelstein coordinates on Schwarzschild's spacetime. Furthermore, we perform the conformal compactification of the physical spacetime with the conformal factor $\Omega = 1/r$. In Section \ref{energy estimates}, we begin by describing the conformally invariant wave equation and then introduce the vector field method. We choose a causal observer, a stress-energy tensor and compute the associated energy current. We add by hand a zero-order term, involving the scalar curvature to the stress-energy tensor for the wave equation, in order to gain an $L^2_{\mathrm{loc}}$ control in our energy current. In the Schwarzschild region, the modified energy current is divergence-free.

The boundary chosen in our framework is composed one the one hand of the future event horizon $\mathscr{H}^+$ and the future null infinity $\mathscr{I}^-$, on the other hand, this is made of the past event horizon $\mathscr{H}^-$ and the past null infinity $\mathscr{I}^-$. Then, Theorem \ref{th_estimates} establishes an equivalence between the energy of the rescaled field on a Cauchy hypersurface and the energy on the future and past boundary of the compactified spacetime. In Section \ref{conformal scattering}, we construct the scattering operator and prove that it is well-defined as an isomorphism between $\mathscr{H}^-\cup \mathscr{I}^-$ and $\mathscr{H}^+\cup \mathscr{I}^+$. We do this by defining energy spaces for initial data and for scattering data in the future on $\mathscr{H}^+\cup \mathscr{I}^+$. Then, we define the future trace operator $\mathcal{T}^+$ that to the initial data defined at $t=0$ associates the trace of the solution on $\mathscr{H}^+\cup \mathscr{I}^+$. Finally, we solve the Goursat problem using the ideas of Hörmander in order to prove that $\mathcal{T}^+$ is an isomorphism. A similar construction holds for the past trace operator $\mathcal{T}^-$.
\section{Geometrical framework}
\label{section:scattering-geometry}
\subsection{The Vaidya spacetime}

The Schwarzschild metric describes a static space-time with a spherical, isolated black hole of constant mass $M$; in spherical coordinates $(t,r,\theta,\phi)$ it is given by :
\begin{equation}
    g_{\mathrm{Sch}} = F(r) \deriv t^2 - F(r)^{-1} \deriv r^2 - r^2 \deriv \omega^2\,, \label{g_sch_t-r}
\end{equation}
with :
\begin{equation*}
    F(r) =1- \dfrac{2M}{r}, \deriv \omega^2 = \deriv \theta^2 + \sin^2\theta \deriv \phi^2 \,.
\end{equation*}
The metric has a curvature singularity at $r=0$. The locus $r=2M$ is a fictitious singularity that can be understood as the union of two null hypersurfaces (the future and the past event horizon) by means of Eddington-Finkelstein coordinates. Outgoing Eddington-Finkelstein coordinates $(u,r,\theta, \phi)$ are defined by $u=t-r_{\star}$, with 
\begin{equation}
   r_{\star} = r + 2M\log\left(r - 2M\right),\hspace{1cm}  \deriv r_{\star} = \dfrac{\deriv r}{F(r)} \,. 
\end{equation}
In these coordinates the Schwarzschild metric reads 
\begin{equation}
    g_{\mathrm{Sch}} = F\deriv u^2 + 2 \deriv u \deriv r - r^2 \deriv \omega\,. \label{Scwharzschild-outgoing}
\end{equation}
The Vaidya metric is defined from \eqref{Scwharzschild-outgoing} by allowing the mass $m$ to depend of the retarded time $u$.

\begin{equation}
    g = F(u,r) \deriv u^2 + 2 \deriv u \deriv r - r^2 \deriv \omega^2, \hspace{0.8cm} F(u,r) = 1 - \dfrac{2m(u)}{r} \label{vaidya-outgoing}
\end{equation}
Alternatively we can construct the Vaidya metric using ingoing Eddington-Finkelstein coordinates $(v,r,\theta, \phi)$ with $v = t + r_{\star}$. 
\begin{equation}
    g = F(v,r) \deriv v^2 - 2 \deriv v \deriv r - r^2 \deriv \omega^2, \hspace{0.8cm} F(v,r) = 1 - \dfrac{2m(v)}{r} \label{vaidya-ingoing}\,,
\end{equation}
Metrics \eqref{vaidya-outgoing} and \eqref{vaidya-ingoing} are solutions of Einstein equations with a source. \eqref{vaidya-outgoing} describes a white hole that evaporates classically via the emission of null dust whereas \eqref{vaidya-ingoing} corresponds to a black hole that mass increases as a result of accretion of null dust.   

In this paper, we will deal with an evaporating white hole, hence the mass $m$ is a decreasing function of $u$. We assume that the mass is a smooth decreasing function of $u$. 

Conformal compactification of Vaidya spacetime, with conformal factor $\Omega = R = 1/r$ :
\begin{equation}
    \hat{g} = R^2 F(u,R) \deriv u^2 - 2 \deriv u \deriv R - \deriv\omega, \hspace{0.8cm} F(u,R) = 1 - 2m(u)R. 
\end{equation}
The inverse rescaled metric is  :
\begin{equation}
    \hat{g}^{-1} = -R^2 F(u,R) \partial_R^2 - 2 \partial_R \partial_u - \partial^2_{\omega^2}\,.
\end{equation}
The d'Alembertien associated to this compactified metric
\[ \square_{\hat{g}} := \nabla_a \nabla^a \, ,\]
where $\nabla$ is the Levi-Civita connection associated to $\hat{g}$, is given in terms of coordinates ($u,R,\omega$) by :
\begin{equation}
    \square_{\hat{g}} = -2\partial_u \partial_R - \partial_R R^2(1-2m(u)R)\partial_R - \Delta_{S^2}\,, \label{d_alembertien_wave-operator}
\end{equation}
The Ricci scalar, is defined by $\hat{g}^{ab}R_{ab}$ i.e. the trace of the Ricci tensor :  
\begin{equation*}
    \mathrm{Scal}_{\hat{g}} = R^{ab}R_{ab} = 12m(u)R \;, ~  \mathrm{Scal}_{g} = 0\,.
\end{equation*}

The metric \eqref{vaidya-outgoing} has the following non-zero Christoffel symbols, with $m'(u) = \deriv m/\deriv u$ :
\begin{gather*}
\Gamma_{ 0 \, 0 }^{ 0  }  =  -3R^2 m(u) + R \, ,~ \Gamma_{ 0 \, 1 }^{ 1  }  =  3R^2 m(u) - R \, , \\
\Gamma_{ 0 \, {0} }^{ \, 1}  =  6R^5 m^2(u) - 5 R^4 m(u) + R^3 m'(u) + R^3  \,   , \\
\Gamma_{ 3 \, 3 }^{ 2 }  =  -\cos\theta\sin\theta \, ,~ \Gamma_{ 3 \, 2 }^{ 3 } =  \frac{\cos \theta}{\sin \theta} \, .
\end{gather*}

\subsection{The second optical function on Vaidya space-time}

We can construct a second optical function analogous to $v= t+ r_{\star}$ in the Schwarzschild spacetime. This has been done in details in \cite{2021_Coudray-Nicolas}. 
On the Schwarzschild metric we have $g = F\deriv u \deriv v - r^2 \deriv \omega^2$; for Vaidya's spacetime we have 
\begin{equation*}
    g = F \deriv u (\deriv u + \dfrac{2}{F}\deriv r) - r^2 \deriv \omega^2\,,
\end{equation*}
and the 1-form $ \deriv u + \dfrac{2}{F}\deriv r$ is not exact. However, introducing an auxiliary positive function $\varphi$ we can write :
\begin{equation}
g = \dfrac{F}{\psi}\deriv u \left(\psi \deriv u +  2 \psi F^{-1} \deriv r\right) \label{metric_varphi} - r^2 \deriv \omega^2\,,
\end{equation}
and arrange for the 1-form, $\psi \deriv u +  2 \psi F^{-1} \deriv r$ to be exact and null, if we assume :
\begin{equation} \label{EDOphi}
    \dfrac{\partial \psi}{\partial u} - \dfrac{F(u,r)}{2}\dfrac{\partial\psi}{\partial r} + \dfrac{2m'(u)}{Fr} \psi = 0 \, .
\end{equation}
This equation can be solved by setting, say, $\psi = 1$ on $\mathscr{I}^{-}$ and integrating along incoming principal null geodesics (see \cite{2021_Coudray-Nicolas}). Then we define $v$ by 
\begin{equation*}
    \deriv v= \psi \, \deriv u + 2 \frac{\psi}{F} \, \deriv r \label{deriv_v}
\end{equation*}

and $v = -\infty$ on $\mathscr{I}^{-}$. 

It is useful to define new radial and time variables such that :
\[
\left\{ \begin{array}{l} {t = u + \tilde{r}\, ,}\\
{t = v - \tilde{r}\, .} \end{array} \right.
\]

The relations between their differentials are 

\begin{align}
      \deriv \Tilde{r} =& \dfrac{1}{2}\left(\psi - 1 \right) \deriv u + \frac{\psi}{F} \deriv r \label{deriv_r_tilde}\\
      \deriv t = & \dfrac{1}{2}\left( \psi + 1 \right) \deriv u + \frac{\psi}{F} \deriv r \label{deriv_t}
\end{align}

\subsection{Variation of the mass and Penrose diagram}

\begin{figure}[!ht]
    \centering
    \includegraphics[scale=0.45]{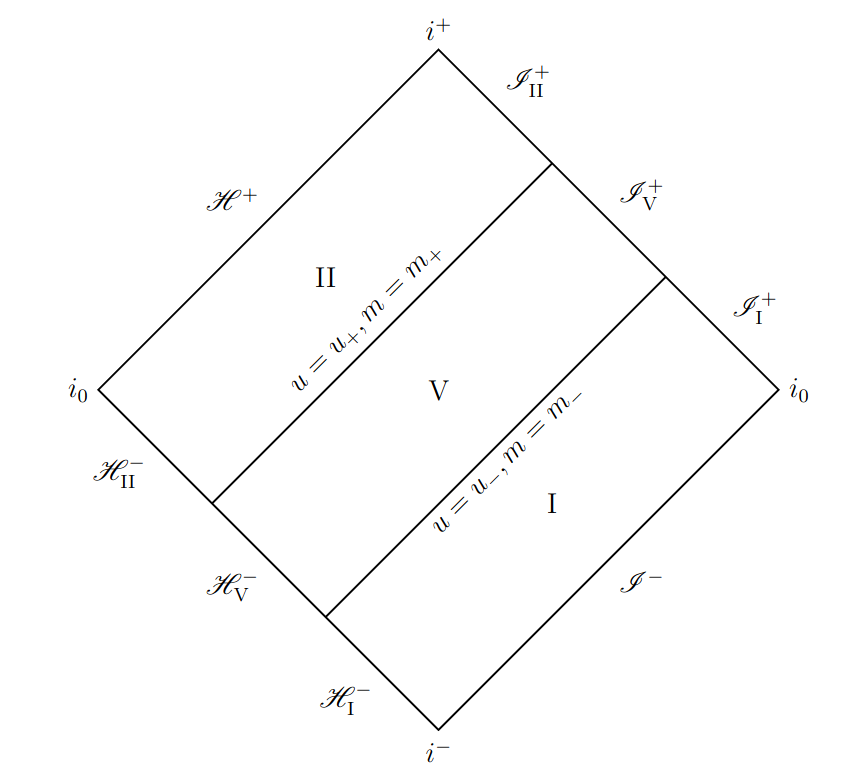}
    \caption{Carter-Penrose's diagram of the exterior of the conformal compactified spacetime. }
    \label{fig:CP_diagram}
\end{figure}

In this work we focus on a Vaidya spacetime that starts from a Schwarzschild spacetime then evolves for a finite retarded time interval towards another Schwarzschild spacetime as we describe in figure \ref{fig:CP_diagram}. The global metric on the compactified spacetime is :
\begin{equation*}
    \hat{g} = R^2 (1-2m(u)R) \deriv u^2 - 2\deriv u \deriv R - \deriv \omega^2\, ,
\end{equation*}
with, for $-\infty<u_-<u_+<+\infty$ given :  
\begin{equation}
    m(u) = \left\lbrace \begin{array}{cl}
        m_{-} & \text{for } u \in \, ]-\infty, u_{-}]\,, \\
        m(u)  & \text{for } u \in [u_{-}, u_{+}]\,,\\
        m_{+} & \text{for } u \in [u_{+}, + \infty[
    \end{array} \right. \, . \label{def_mass}
\end{equation}
where $m_{+} < m_{-}$.

Boundaries between Vaidya and Schwarzschild areas are two $u=\cte$-hypersurfaces : $S_{u_+}$ and $S_{u_-}$ for respectively $\lbrace u=u_{+} \rbrace$ and $\lbrace u=u_{-} \rbrace$. We denote by $\mathrm{I}$ and $\mathrm{II}$ respectively the past and future Schwarzschild spacetimes and by $\mathrm{V}$ the Vaidya domain in between. We have also $\mathscr{H}^{-} =  \mathscr{H}^{-}_{\mathrm{I}} \cup \mathscr{H}^{-}_{\mathrm{V}} \cup  \mathscr{H}^{-}_{\mathrm{II}}$ (see remark \ref{horizon}) with :
\begin{align*}
    \mathscr{H}^{-}_{\mathrm{I}} =& \mathscr{H}^{-}\cap\lbrace u\leq u_{-}\rbrace \,,\\
    \mathscr{H}^{-}_{\mathrm{V}} =& \mathscr{H}^{-}\cap \lbrace u_{-} \leq u\leq u_{+}\rbrace\,, \\
    \mathscr{H}^{-}_{\mathrm{II}} =&\mathscr{H}^{-}\cap \lbrace u\geq u_{+}\rbrace \, ,
\end{align*}
and $\mathscr{I}^{+} = \mathscr{I}^{+}_{\mathrm{I}} \cup \mathscr{I}^{+}_{\mathrm{V}} \cup \mathscr{I}^{+}_{\mathrm{II}}$ with : 
\begin{align*}
    \mathscr{I}^{+}_{\mathrm{I}} =& \mathscr{I}^{+}\cap\lbrace u\leq u_{-}\rbrace\,, \\
    \mathscr{I}^{+}_{\mathrm{V}} =& \mathscr{I}^{+}\cap \lbrace u_{-} \leq u\leq u_{+}\rbrace\,, \\
    \mathscr{I}^{+}_{\mathrm{II}} =&\mathscr{I}^{+ }\cap \lbrace u\geq u_{+}\rbrace \,.
\end{align*}
In the remainder of this article, we shall denote by $\Sigma_{t}$ the level hypersurfaces of $t$, in particular :
\begin{align*}
    \Sigma_0 =& \lbrace (t,r,\omega) \vert t=0, r\in [r_h, +\infty[, \omega \in S^2\rbrace\,, \\
    \Sigma_0^V =& \Sigma_0 \cap \lbrace u_-\leq u \leq u_+\rbrace \, . 
\end{align*}
where $r_h$ refers to the past horizon (see Remark \ref{horizon}). 
\begin{remark}
\label{horizon}
The construction of the past horizon $\lbrace r = r_h(u)\rbrace$ can be found in \cite{2021_Coudray-Nicolas}, where the authors study its behaviour in general and devote particular attention to the case where the transition happens in finite retarded time. The function $r_h(u)$ satisfies :
\begin{enumerate}
    \item $r_h(u) = 2m_-$, for $u\leq u_-$. 
    \item $r_h(u)$ is decreasing.
    \item $\lim_{u\to +\infty}r_h(u) = 2m_+$.
\end{enumerate}
Generically, $r_h(u)>2m_+$ on $[u_+,+\infty[$. So, although $\mathcal{H}_{\mathrm{I}}^-$ coincides with the past Schwarzschild horizon in the region I, the same is not in general true of $\mathcal{H}_{\mathrm{II}}^-$ and region II.
\end{remark}
\section{Energy estimates}
\label{energy estimates}
\subsection{Strategy and Propositions}
Let $\Psi$ be a solution of the physical wave equation on the Schwarzschild spacetime :
\begin{equation}
    \left(\square_{g} + \dfrac{1}{6}\mathrm{Scal}_g\right) \Psi = \square_g \Psi = 0, \label{eq_ondes_physical}
\end{equation}
with $\square_g = \tilde{\nabla}^a\tilde{\nabla}_a$ that is the d'Alembertian operator associated to physical metric $g$, with the Levi-Civita connection $\tilde{\nabla}$. This is equivalent to saying that $\phi = \Omega^{-1} \Psi$ satisfies the conformal wave equation on the rescaled spacetime : 
\begin{equation}
    \left(\square_{\hat{g}} + \dfrac{1}{6}\mathrm{Scal}_{\hat{g}}\right)\phi = 0. \label{eq_ondes_rescaled}
\end{equation}

We will study the scattering for \eqref{eq_ondes_physical} at the level of the conformal field $\phi$ by computing energy estimates for the solutions to \eqref{eq_ondes_rescaled}. This can be done by choosing a stress-energy tensor $T_{ab}$ and a causal vector field $V^a$ and contracting these two quantities to obtain an energy current : 
\begin{equation}
    J_a = V^b T_{ab}\,. 
\end{equation}
Finally we define the energy flux $\mathcal{E}_{\mathcal{S}}$ measured by the observer $V$ across a hypersurface $\mathcal{S}$ by : 
\begin{equation*}
    \mathcal{E}_{V,\mathcal{S}}(\phi) = \int_{\mathcal{S}} \star T_{ab}V^b \deriv x^b = \int_{\mathcal{S}} T_{ab} V^a n^b l \lrcorner\dvol\, ,
\end{equation*}
where $\star$ denotes the Hodge dual, $l$ is a vector field transverse to the hypersurface $\mathcal{S}$ and $n$ is the normal vector to $\mathcal{S}$ such that $\hat{g}(l,n) = 1$. The energy momentum tensor for the rescaled field is taken as :
\begin{equation}
    T_{ab} = \nabla_a \phi \nabla_b \phi - \dfrac{1}{2}\hat{g}_{ab} \nabla_c \phi \nabla^c\phi + \frac{1}{12}\mathrm{Scal}_{\hat{g}}\phi^2 \hat{g}_{ab}\, .  \label{energy-tensor}
\end{equation}
The divergence of the stress-energy tensor in the conformal spacetime $(\hat{\mathcal{M}}, \hat{g})$ is not zero since : 
\begin{equation*}
    \nabla^a T_{ab} = \square_{\hat{g}}\phi \nabla_b \phi + \frac{1}{12}\nabla_b \left( \mathrm{Scal}_{\hat{g}} \phi^2 \right) \, .
\end{equation*}
Contracting this divergence with the timelike observer $V^b = \partial_u^b$ and using the wave equation \eqref{eq_ondes_rescaled} to replace $\square_{\hat{g}}$, we obtain : 
\begin{equation*}
    V^b \nabla^a T_{ab} = -\frac{1}{6}\mathrm{Scal}_{\hat{g}}\,  \phi\,  \phi_u + \frac{1}{12}\partial_u\left(\mathrm{Scal}_{\hat{g}} \phi^2\right)\, . 
\end{equation*}
We observe that this is zero if the scalar curvature $\mathrm{Scal}_{\hat{g}}$ does not depend on $u$. It happens in the Schwarzschild area where consequently : 
\begin{equation}
     V^b \nabla^a T_{ab}\vert_{\mathrm{Sch}} = 0\, , \label{div_tab_sch}
\end{equation}
However, in the Vaidya region, $\mathrm{Scal}_ {\hat{g}} = 12m(u)R$ and then : 
\begin{equation*}
     V^b \nabla^a T_{ab}\vert_{\mathrm{V}} = m'(u) R \phi^2\, . 
\end{equation*}
\begin{remark}
    
Let $\Sigma_0$ be the hypersurface of initial data at $t=0$. In this paper, we adopt the following strategy (introduced in \cite{2016_Nicolas}) : we take initial data $(\phi\vert_{t=0}, \partial_t \phi\vert_{t=0})\in \mathcal{C}^{\infty}_{0}(\Sigma_0)\times \mathcal{C}^{\infty}_{0}(\Sigma_0)$ and the associated solution $\phi$ of \eqref{eq_ondes_rescaled}. Then we obtain estimates between the energy flux of initial data (denoted $\mathcal{E}_{\Sigma_0}(\phi)$) and the energy flux of the solution at the boundary of the conformal spacetime. Estimates proved for data in $(\phi\vert_{t=0}, \partial_t \phi\vert_{t=0})\in \mathcal{C}^{\infty}_{0}(\Sigma_0)\times \mathcal{C}^{\infty}_{0}(\Sigma_0)$ are then extended by density for any data with finite energy. Finally, the trace operator acts from $\mathcal{H}_0$, the energy space of initial data $(\phi\vert_{t=0}, \partial_t \phi\vert_{t=0})\in \mathcal{C}^{\infty}_{0}(\Sigma_0)\times \mathcal{C}^{\infty}_{0}(\Sigma_0)$, completed in the norm $\left[\mathcal{E}_{\Sigma_0}(\phi)\right]^{1/2}$ to the energy space of the trace of the solution $\phi$ at the boundary (see definition \ref{def_Ht} and \ref{def-H+} for more details). 

\end{remark}

Our approach to the scattering is based on energy estimates. By applying Stokes' Theorem, we can derive equalities and inequalities between the energy fluxes through our different hypersurfaces.  It is important to notice that on the Schwarzschild spacetime (see section \ref{EE-Schwarzschild}) there exists an exact conservation law coming, firstly from \eqref{div_tab_sch} where the contraction of the Killing observer $\partial_u$ and the divergence of the stress-energy tensor is zero, and secondly from the Killing equation that constrains $\nabla^{(a}V^{b)}$ to be zero also.  This is not true on the Vaidya spacetime (see section \ref{EE-Vaidya}) where we have only an approximate conservation law. This leads to the first theorem of this article, which establishes an equivalence between the energy flux at $t=0$ and the energy fluxes on the boundary.
\begin{theorem}   
\label{th_estimates}
For $(\phi\vert_{t=0}, \partial_t\phi\vert_{t=0}) \in \mathcal{C}^{\infty}_{0}(\Sigma_0) \times \mathcal{C}^{\infty}_{0}(\Sigma_0)$ we have : 
\begin{align*}
      \mathcal{E}_{\Sigma_0} (\phi) & \simeq \mathcal{E}_{\mathscr{H}^{+}}(\phi) + \mathcal{E}_{\mathscr{I}^{+}}(\phi) \, , \\
      \mathcal{E}_{\Sigma_0} (\phi) & \simeq \mathcal{E}_{\mathscr{I}^{-}}(\phi) + \mathcal{E}_{\mathscr{H}^{-}}(\phi) \, . 
    \end{align*}
\end{theorem}
The proof of this theorem is decomposed in two parts : firstly we will prove that on the Schwarzschild spacetime we have : 
\begin{proposition}
\label{prop_ee_s}
The global energy conservation laws on the Schwarzschild spacetime  (see figures \ref{fig:initial_sch} and \ref{fig:boundary_sch}):
    \begin{align}
    \mathcal{E}_{\mathscr{H}^{-}_{\mathrm{I}}} (\phi) + \mathcal{E}_{\mathscr{I}^{-}}(\phi) &= \mathcal{E}_{\mathcal{S}_{u_-}}(\phi) + \mathcal{E}_{\mathscr{I}^{+}_{\mathrm{I}}}(\phi)\label{estimate_I} \, ,\\
    \mathcal{E}_{\mathscr{H}^{+}} (\phi) + \mathcal{E}_{\mathscr{I}^{+}_{\mathrm{II}}}(\phi) &= \mathcal{E}_{\mathcal{S}_{u_+}}(\phi) + \mathcal{E}_{\mathscr{H}^{-}_{\mathrm{II}}}(\phi) \label{estimate_II}\, .
\end{align}
can be decomposed in region II into two conservation laws between : 
\begin{enumerate}
    \item the hypersurface $\tilde{\Sigma}_0^{II}$ defined by : 
    \begin{equation*}
        \tilde{\Sigma}_0^{II} = \left(\Sigma_0\cap\lbrace u\geq u_+\rbrace\right) \cup \left(\mathcal{S}_{u_+}\cap \lbrace t\geq 0\rbrace\right)\, 
    \end{equation*}
    \item the future and the past boundary of the II-region referred to respectively as $\mathcal{B}_+^{II}$ and $\mathcal{B}_-^{II}$ :
    \begin{align*}
        \mathcal{B}_+^{II} = & \mathscr{H}^+ \cup \mathscr{I}_{II}^+\, ,\\
        \mathcal{B}_-^{II} =& \mathscr{H}^-_{II} \cup \left( \mathcal{S}_{u_+}\cap\lbrace t\leq 0\rbrace\right)\, . 
    \end{align*}
\begin{figure}[!ht]
    \begin{minipage}[c]{.46\linewidth}
        \centering
        \includegraphics[width=9cm]{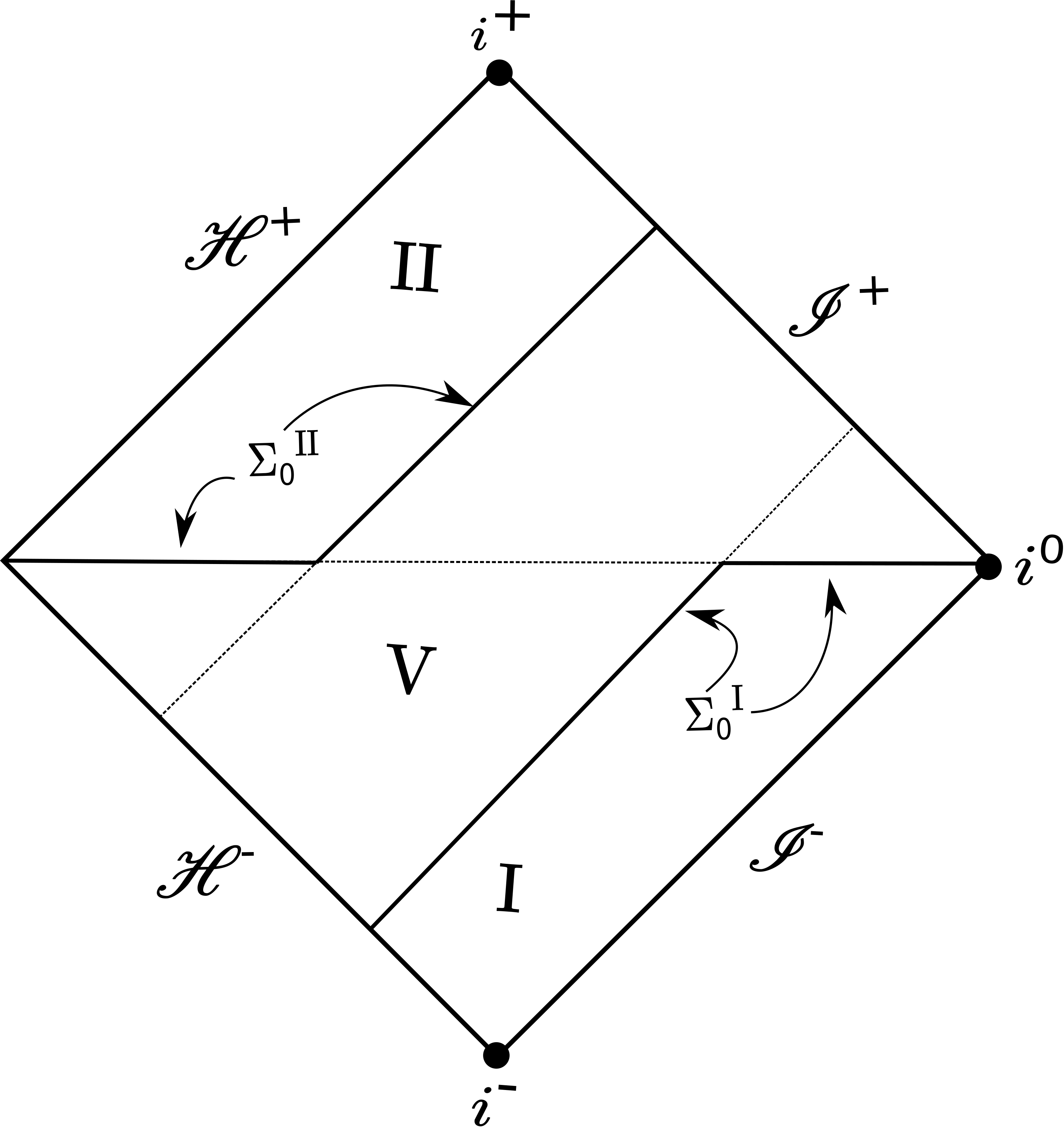}
        \caption{Hypersurfaces of "initial data" ($\Sigma_0^I$ and $\Sigma_0^{II}$) on the Schwarzschild area.}
        \label{fig:initial_sch}
    \end{minipage}
    \hfill%
    \begin{minipage}[c]{.46\linewidth}
        \centering
        \includegraphics[width=9cm]{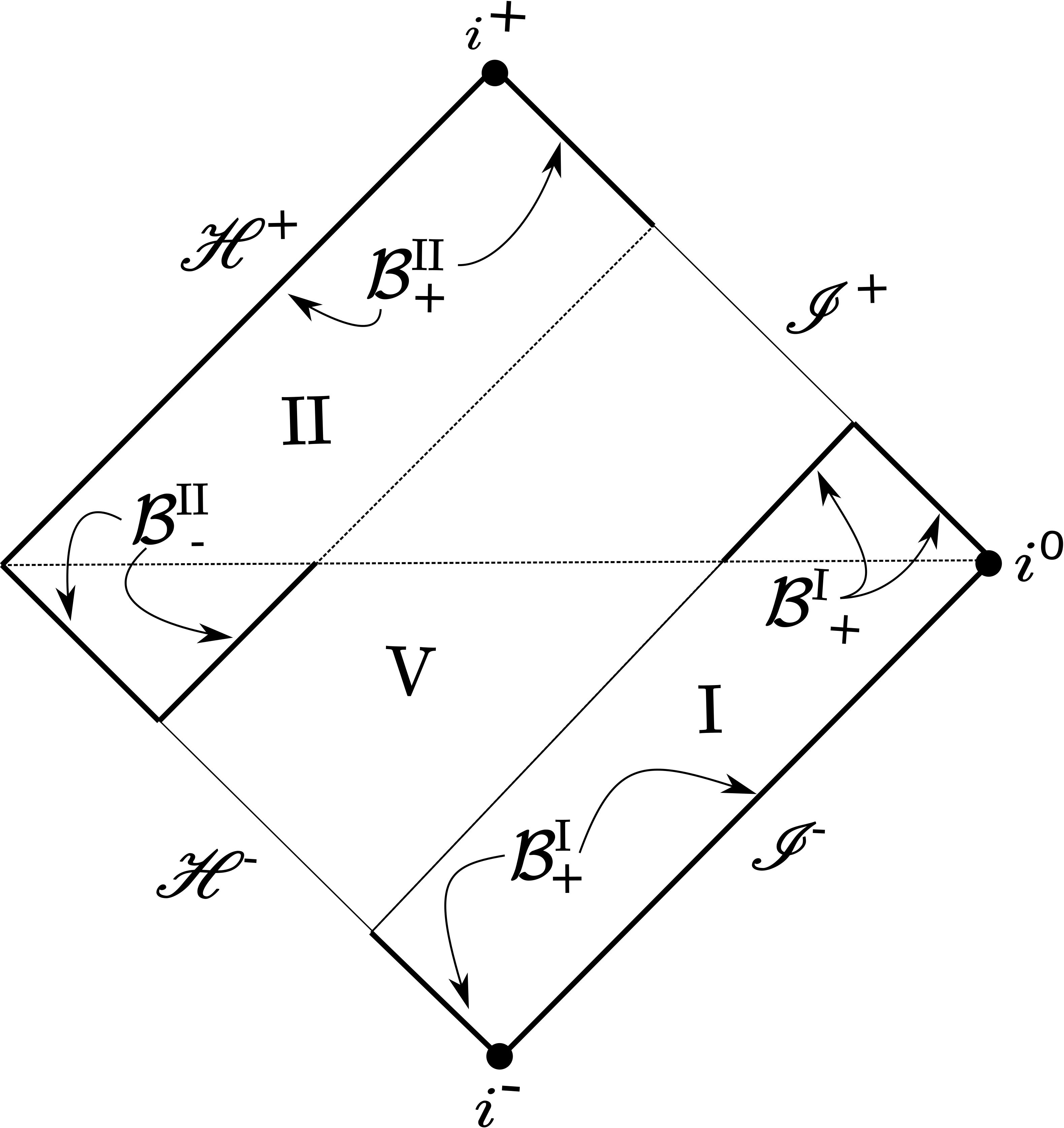}
        \caption{Past and future boundary on the Schwarzschild zone.}
        \label{fig:boundary_sch}
    \end{minipage}
\end{figure}
\end{enumerate}    
This leads to :
\begin{align}
    \mathcal{E}_{\tilde{\Sigma}_0^{\mathrm{II}}}(\phi)  &=  \mathcal{E}_{\mathcal{B}_+^{\mathrm{II}}}(\phi)\, , \\
     \mathcal{E}_{{\Sigma}_0\cap \lbrace u \geq u_+\rbrace}(\phi) & =  \mathcal{E}_{\mathcal{B}_-^{\mathrm{II}}}(\phi)\, .
\end{align}
The same decomposition holds in the I-region between this time :
\begin{enumerate}
    \item $\tilde{\Sigma}_0^\mathrm{I}$, the hypersurface : 
    \begin{equation*}
        \tilde{\Sigma}_0^{\mathrm{I}} = \left(\Sigma_0\cap\lbrace u\leq u_-\rbrace\right) \cup \left(\mathcal{S}_{u_-}\cap \lbrace t\leq 0\rbrace\right)\, 
    \end{equation*}
    \item $\mathcal{B}_+^{\mathrm{I}}$ and $\mathcal{B}_-^{\mathrm{I}}$ the future and the past boundary on the I-region : 
     \begin{align*}
        \mathcal{B}_+^{\mathrm{I}} = & \mathscr{I}^+_{\mathrm{I}} \cup \left(\mathcal{S}_{u_-} \cap\lbrace t\geq 0\rbrace\right)\, , \\
        \mathcal{B}_-^{\mathrm{I}} =& \mathscr{H}^-_{\mathrm{I}} \cup \mathscr{I}^- \,. 
    \end{align*}
\end{enumerate}
and the conservation laws are : 
\begin{align}
    \mathcal{E}_{\Sigma_0\cap \lbrace u \leq u_-\rbrace}(\phi) &=  \mathcal{E}_{\mathcal{B_+^{\mathrm{I}}}}(\phi)\, , \\
     \mathcal{E}_{\tilde{\Sigma}_0^{\mathrm{I}}}(\phi) & =  \mathcal{E}_{\mathcal{B_-^{\mathrm{I}}}}(\phi)\, .
\end{align}
\end{proposition}
\noindent Then, we will focus our attention on the Vaidya spacetime where we don't have conservation laws but only approximate conservation laws. 
\begin{proposition} \label{prop_ee_v}
The global approximate conservation law in the Vaidya area is given by : 
    \begin{equation*}
        \mathcal{E}_{\mathscr{H}^{-}_{\mathrm{V}}}(\phi) + \mathcal{E}_{\mathcal{S}_{u_{-}}}(\phi) \simeq \mathcal{E}_{\mathcal{S}_{u_{+}}}(\phi) + \mathcal{E}_{\mathscr{I}^{+}_{V}}(\phi) \, . 
    \end{equation*}
This equivalence can be decomposed into two equivalences in the following framework (see figure \ref{fig:geometry_vaidya_past} and \ref{fig:geometry_vaidya_future}) :
    \begin{figure}[!ht]
    \begin{minipage}[l]{.46\linewidth}
        \centering
        \includegraphics[width=9cm]{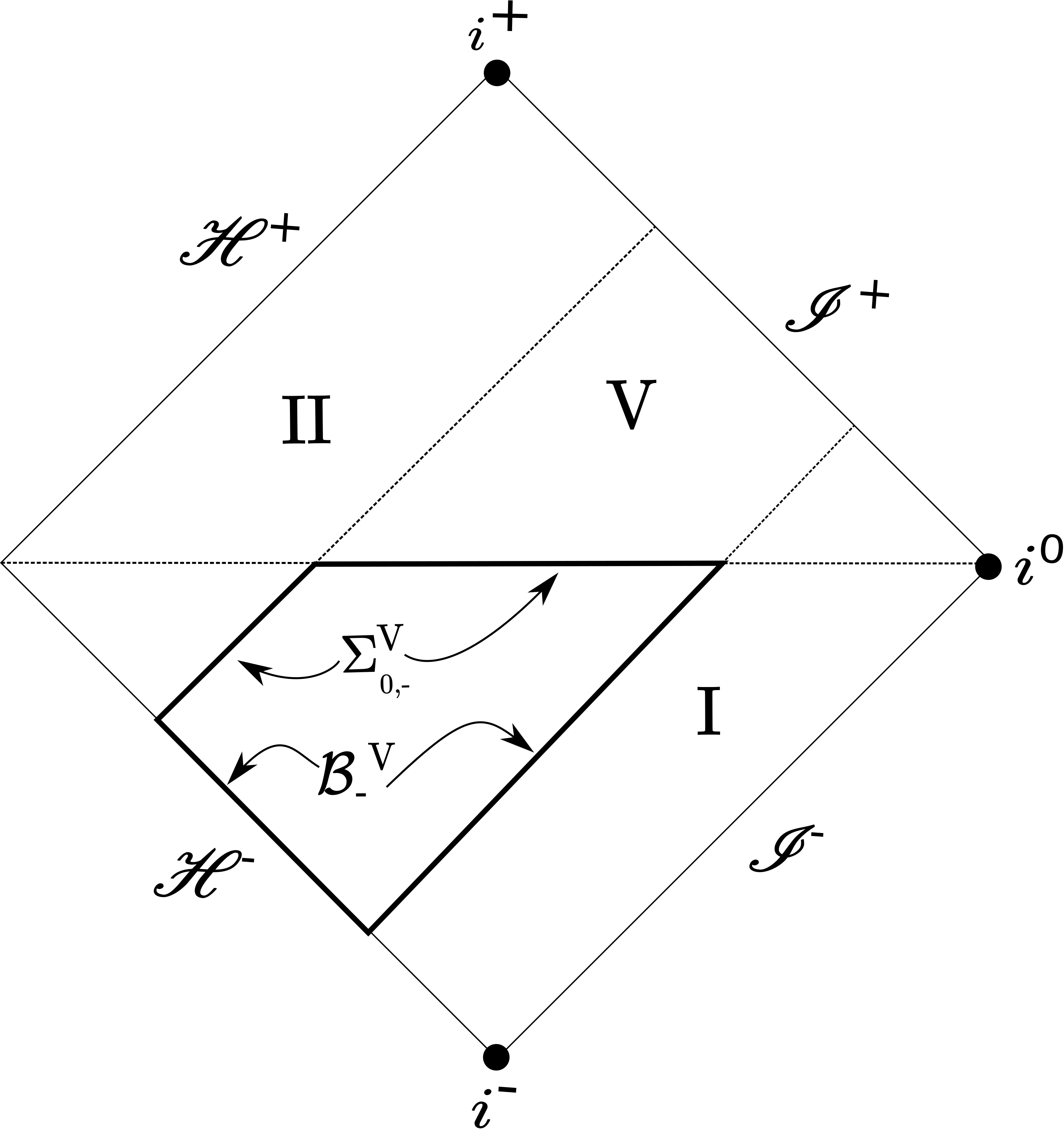}
        \caption{Framework in the past on the Vaidya spacetime.}
        \label{fig:geometry_vaidya_past}
    \end{minipage}
    \hfill%
    \begin{minipage}[c]{.46\linewidth}
        \centering
        \includegraphics[width=9cm]{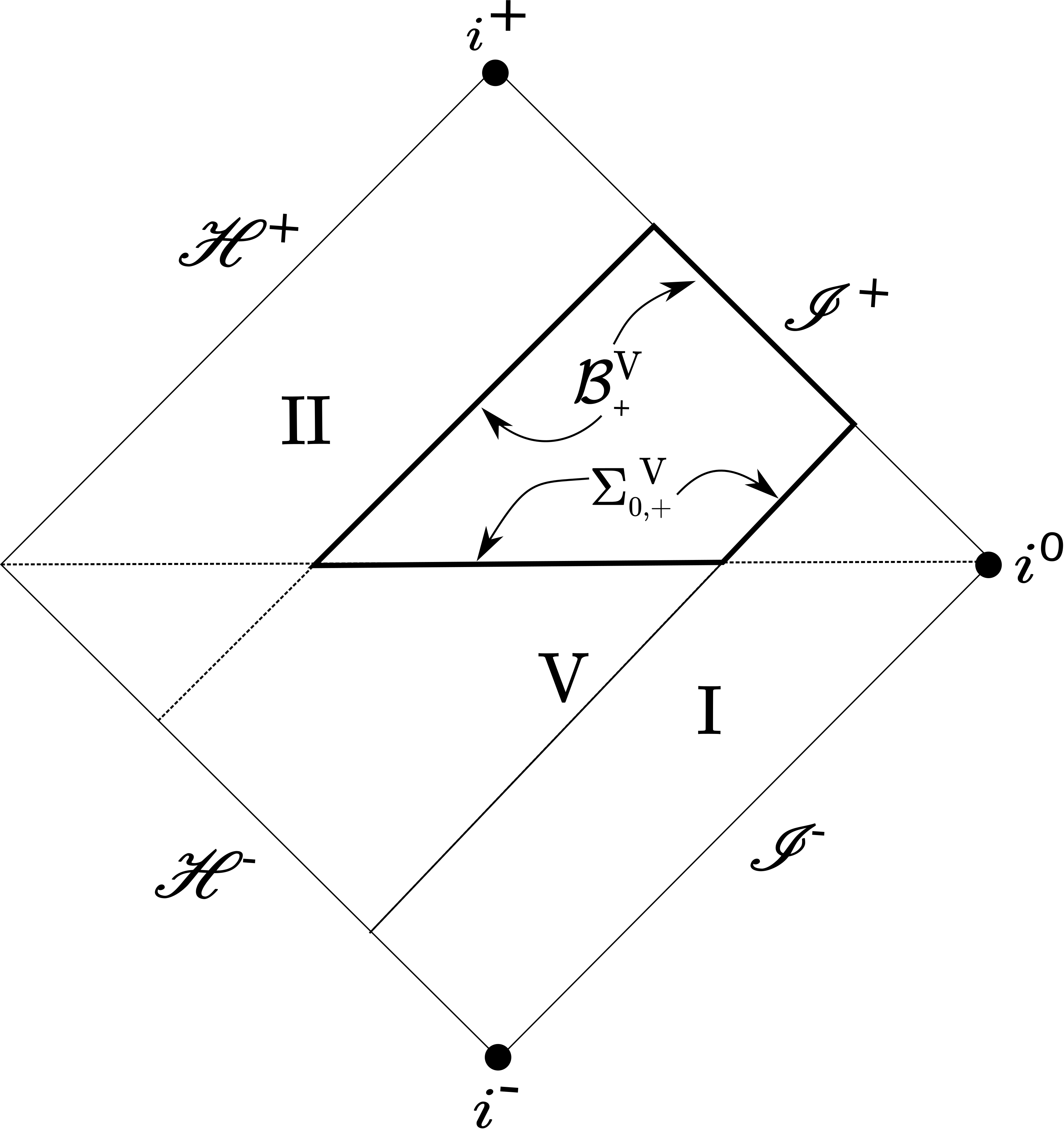}
        \caption{Future boundary on the Vaidya spacetime.}
        \label{fig:geometry_vaidya_future}
    \end{minipage}
\end{figure}
\begin{enumerate}
    \item $\tilde{\Sigma}_{0,+}^{\mathrm{V}}$ and $\tilde{\Sigma}_{0,-}^{\mathrm{V}}$ are on V, the hypersurfaces : 
    \begin{align*}
        \Tilde{\Sigma}_{0,-}^{\mathrm{V}} =& \left(\mathcal{S}_{u_+}\cap \lbrace t\leq 0\rbrace\right) \cup \left( \Sigma_0 \cap \lbrace u\geq u_- \lbrace\right)\, , \\
        \Tilde{\Sigma}_{0,+}^{\mathrm{V}} =& \left(\Sigma_0\cap \lbrace u \leq u_+\rbrace \right)\cup \left( \mathcal{S}_{u_-}\cap \lbrace t \geq 0 \rbrace\right) \, .
    \end{align*}
    \item  the past and future boundary (resp. $\mathcal{B}_-^{\mathrm{V}}$ and $\mathcal{B}_+^{\mathrm{V}}$) are : 
    \begin{align*}
        \mathcal{B}_-^{\mathrm{V}} = & \mathscr{H}^+_V \cup \left(\mathcal{S}_{u_-}\cap \lbrace t \leq 0\rbrace \right)\, , \\
        \mathcal{B}_+^{\mathrm{V}} = & \mathscr{I}^+_V \cup \left(\mathcal{S}_{u_+}\cap \lbrace t\geq 0 \rbrace \right)\, . 
    \end{align*}

\end{enumerate}
The conservation law is divided into two equivalences : 
    \begin{align}
    \mathcal{E}_{\tilde{\Sigma}_{0,-}^{\mathrm{V}}}(\phi)  &\simeq \mathcal{E}_{\mathcal{B}_{-}^{\mathrm{V}}}(\phi) \, , \label{approx-vaidya-past}\\
     \mathcal{E}_{\tilde{\Sigma}_{0,+}^{\mathrm{V}}}(\phi) & \simeq  \mathcal{E}_{\mathcal{B}_{u_+}^{\mathrm{V}}}(\phi)\, . \label{aprox-vaidya-future}
\end{align}
\end{proposition}
\textbf{Proof :} The proof is done in the two following sections, respectively section \ref{EE-Schwarzschild} for Proposition \ref{prop_ee_s} and section \ref{EE-Vaidya} for Proposition \ref{prop_ee_v}.

\subsection{Energy estimates on the Schwarzschild spacetime}
\label{EE-Schwarzschild}
Energy estimates on the Schwarzschild spacetime have been obtained in \cite{2016_Nicolas}, and we adapt these results to our framework. We choose the Killing vector field $T^a = \partial_u^a$ as the observer on the Schwarzschild spacetime. Due to $\hat{g}(V,V) = R^2 F$, it is clearly timelike and it is furthermore future oriented on the rescaled spacetime.
The expression of energy fluxes across the hypersurfaces $\Sigma_t$, $\mathscr{I}$, $\mathscr{H}^{-}$ and $\mathcal{S}_u$  are as follows : first, the energy current is given by : 
\begin{equation}
    \star V^a T_{ab} = \left[\phi_u^2 + R^2F \phi_u \phi_R \right]\deriv u \wedge \deriv \omega + \left[ m(u)R^2 \phi^2 + \frac{R^2F}{2}\phi_R^2 + \frac{1}{2} \vert \nabla_{S^2}\phi\vert^2 \right] \deriv R\wedge \deriv \omega\, , \label{courant}
\end{equation}
on $\Sigma_t$, 
\begin{equation*}
    \deriv u =  \frac{\deriv R}{R^2 F}\, ,
\end{equation*}
hence,  %
\begin{align}
    \mathcal{E}_{\Sigma_t}(\phi) &= \int_{\Sigma_t} \left[ \phi_u^2  + R^2 F \phi_u \phi_R + \frac{R^2 F}{2}\left( 2m(u)R^2 \phi^2 + R^2 F\phi_R^2 +  \vert \nabla_{S^2}\phi\vert^2 \right) \right] \deriv u \wedge \deriv\omega \, , \label{energie_sigma_t} 
    \intertext{and,}
    \mathcal{E}_{\mathscr{I}^{+}}(\phi) &= \int_{\mathscr{I}^{+}} \phi_u^{2}\, \deriv u \wedge\deriv \omega \, ,\label{energie_scri} \\
    \mathcal{E}_{\mathcal{S}_{u_{\pm}}}(\phi) &= \int_{\mathcal{S}_{u_{\pm}}}\left( m_{\pm}R^2 \phi^2 + \frac{R^2(1-2m_{\pm}R)}{2}\phi_R^2 + \frac{1}{2} \vert \nabla_{S^2}\phi\vert^2 \right) \deriv R\wedge \deriv \omega \, . \label{energie_Su}\\
\end{align}
Knowing that on the past horizon we have (see \cite{2021_Coudray-Nicolas}) :
\begin{equation}
    \deriv R = \frac{R^2 F}{2}\deriv u \, , 
\end{equation}
then :
\begin{equation*}
    \mathcal{E}_{\mathscr{H}^{-}}(\phi) = \int_{\mathscr{H}^-} \left[\phi_u^2 + \frac{R_h^2F}{2} \left(2\phi_u \phi_R + 2m(u) R_h^2 F \phi^2 + R_h^2F\phi_R^2 + \vert \nabla_{S^2}\phi\vert^2 \right)\right]\deriv u \wedge \deriv \omega\, , 
\end{equation*}
with $R_h= \frac{1}{r_h(u)}$. In the region $\mathrm{I}$, on the horizon, $F = 1 - 2m_-R$ is zero, then we decompose the previous expression into : 
\begin{align*}
        \mathcal{E}_{\mathscr{H}^{-}}(\phi) = & \int_{\mathscr{H}_{\mathrm{I}^-} }\phi_u^2 \deriv u \wedge \deriv \omega  \\
        &\, + \int_{\mathscr{H}^-_{\mathrm{V}}\cup \mathscr{H}^-_{\mathrm{II}}}\left[\phi_u^2 + \frac{R_h^2F}{2} \left(2\phi_u \phi_R + 2m(u) R_h^2 F \phi^2 + R_h^2F\phi_R^2 + \vert \nabla_{S^2}\phi\vert^2 \right)\right]\deriv u \wedge \deriv \omega\, . 
\end{align*}

The coordinates $(u,r,\theta, \varphi)$ are unsuitable to compute the energy fluxes through $\mathscr{H}^{+}$ and $\mathscr{I}^{-}$ because these hypersurfaces are such that $u=\pm \infty$ and $R=\cte$ on them. Since $\mathscr{H}^+$ and $\mathscr{I}^-$ are in Schwarzschild regions (respectively $\mathrm{II}$ and $\mathrm{I}$), there is no difficulty to turn to $(v,R,\theta, \varphi)$ coordinates, with $v = t - r_{\star}$ and $\partial_t^a = \partial_v^a$, then : 
\begin{equation}
    \star V^a T_{ab} = -\left[\phi_v^2 - R^2F \phi_v \phi_R \right]\deriv v \wedge \deriv \omega - \left[ \frac{1}{12}\mathrm{Scal}_{\hat{g}} \phi^2 + \frac{R^2F}{2}\phi_R^2 + \frac{1}{2} \vert \nabla_{S^2}\phi\vert^2 \right] \deriv R\wedge \deriv \omega\, . \label{courant_v}
\end{equation}
Hence : 
\begin{align}
    \mathcal{E}_{\mathscr{I}^{-}}(\phi) &= \int_{\mathscr{I}^{-}} \phi_v^2 \deriv v\wedge\deriv \omega\, , \\
    \mathcal{E}_{\mathscr{H}^{+}}(\phi) &= \int_{\mathscr{H}^+} \phi_v^2 \deriv v\wedge \deriv \omega\, . 
\end{align}
The energy flux $\mathcal{E}_{\Sigma_t}(\phi)$ is equivalent to : 
\begin{equation*}
     \mathcal{E}_{\Sigma_t}(\phi) \simeq \int_{\Sigma_t} \left( \phi_u^2 + R^4 \phi_R^2 + R^3 \phi^2 + R^2 \vert \nabla_{S^2}\phi \vert^2 \right)\deriv u \wedge \deriv \omega
\end{equation*}

On the rescaled Schwarzschild spacetime, $V^a$ is a Killing vector field, i.e. $\nabla^{(a}V^{b)} = 0$ and the stress-energy tensor satisfies \eqref{div_tab_sch}. This ensures that the divergence of the current $J_a$ is zero. 
\begin{equation*}
    \nabla^a J_a = \nabla^{(a} \left(V^{b)}T_{ab}\right) = \nabla^{(a} V^{b)} T_{ab} + V^b \nabla^a T_{ab} = 0\, .
\end{equation*}
From this we infer the conservation of the energy between the hypersurfaces of our boundary and we prove  Proposition \ref{prop_ee_s}. \qed
\begin{remark}
Here, we have neglected the fact that the boundary is not compact between the past (future) horizon and past (future) null infinity. Due to the singularities at $i^-$ and $i^+$, we cannot directly state \eqref{estimate_I} and \eqref{estimate_II} without proving that the energy tends to zero at these singularities. Since we assume that the spacetime is the Schwarzschild spacetime near these two singularities, Nicolas demonstrated in \cite{2016_Nicolas}, relying on results from Dafermos and Rodnianski (refer to Theorem 4.1 in \cite{2008_DAfermos_Rodnianski}), that there is no energy of the rescaled field going to $i^+$ and $i^-$. 

\end{remark}

\subsection{Energy fluxes on the Vaidya spacetime}
\label{EE-Vaidya}
In order to perform energy estimates on the compact area of the Vaidya spacetime, we need to use suitable hypersurfaces. The time function $t$ is problematic because $t\in \mathbb{R}$ is unbounded on $V$. This implies that if we choose the level hypersurfaces of $t$  in our vector field method, we would have to use the Grönwall lemma on an infinite domain. A better idea is to use a family of spacelike hypersurfaces that are transverse to null infinity and the horizon. 
\begin{figure}[!ht]
    \centering
    \includegraphics[scale=0.7]{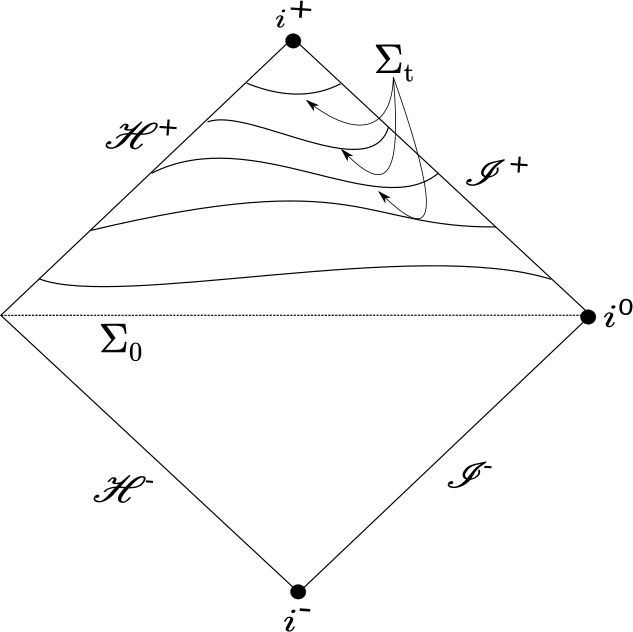}
    \caption{Illustration of the foliation of the rescaled spacetime in the future of $\Sigma_0$ by spacelike hypersurfaces, transverse to $\mathscr{I}^+$ and $\mathscr{H}^+$.}
\end{figure}

\noindent We denoted by $\Sigma_{\tau}$ theses hypersurfaces and we set : 
\begin{equation*}
    \Sigma_{\tau = 0} = \Sigma_{t=0}\,.
\end{equation*}
We introduce the Newman-Penrose tetrad of null vectors on the compactified spacetime $(\hat{l}, \hat{n}, \hat{m}, \hat{\Bar{m}})$ : 
\begin{align}
    \hat{l}^a =& \partial_u^a + \frac{R^2F}{2}\partial_R^a\, ,\\
    \hat{n}^a =& - \partial_R^a\, ,\\
    \hat{m}^a =& \frac{1}{\sqrt{2}}\left(\partial_{\theta}^a + \frac{i}{\sin{\theta}}\partial_{\varphi}^a\right)\, , \hat{\bar{m}}^a =\frac{1}{\sqrt{2}}\left(\partial_{\theta}^a - \frac{i}{\sin{\theta}}\partial_{\varphi}^a\right)\,\, .
\end{align}
It is normalised, i.e. 
\begin{equation*}
    \hat{l}_a\hat{n}^a = - \hat{m}^a\hat{\bar{m}}_a = 1\,,
\end{equation*}
and the spin coefficients associated to this tetrad are\footnote{For the sake of simplicity we denote these spin-coefficient without a hat, however they are the spin coefficient associated with the compactified metric $\hat{g}$ and the associated connection $\hat{\nabla}$. We also denote without a hat directional derivatives $D,\Delta, \delta$ and $\delta'$. } : 
\begin{align*}
        \varepsilon &= \frac{1}{2}R - \frac{3}{2}R^2m(u)\, , \alpha = -\beta = -\frac{\sqrt{2}}{4}\cot{\theta}\, , \\
        \kappa &= \rho = \gamma = \tau = \sigma = \nu = \lambda = \mu = \pi = 0\, . 
\end{align*}
We define $\nu$ as :
\begin{equation}
    \nu = \frac{1}{\sqrt{2}}\left(\hat{n} + \hat{l}\right)\, ,
\end{equation}
It is straightforward that $\nu$ is timelike and  :
\begin{equation*}    
\hat{g}(\nu, \nu) = 1\, . 
\end{equation*}
We use $\nu$ to split the 4-volume measure into : 
\begin{equation*}
    \dvol = (\nu_a \deriv x^a ) \wedge \deriv \Sigma_{\tau}\, . 
\end{equation*}
If $\nu$ is orthogonal to the hypersurfaces $\Sigma_{\tau}$, we have $(\nu_a \deriv x^a) = \deriv \tau$. In other situations, we denote by $k$ the orthogonal vector to the hypersurface $\Sigma_{\tau}$.
The observer $\partial_u$ reads  : 
\begin{equation}
    \partial_u^a = \hat{l}^a + \frac{R^2F(u,R)}{2}\hat{n}^a\, .
\end{equation}
and the energy current is given by  : 
\begin{equation}\label{current_v}
    J_a = T_{ab}\partial_u^a = \left(\phi_{\hat{l}} + \frac{R^2F}{2}\phi_{\hat{n}}\right)\nabla_b \phi 
     + \left(\hat{l}_b + \frac{R^2F}{2}\hat{n}_b\right)\left(\frac{1}{12}\mathrm{Scal}_{\hat{g}} \phi^2- \frac{1}{2}\langle \nabla \phi, \nabla \phi \rangle  \right)\, . 
\end{equation}
The energy flux measured by this observer through $\Sigma_{\tau}$ is now : 
\begin{equation}
    \mathcal{E}_{\partial_u, \Sigma_{\tau}}(\phi) =  \int_{\Sigma_{\tau}} J_a \nu^a \deriv \Sigma_{\tau} = \int_{\Sigma_{\tau}} T_{ab}\partial_u^a \nu^b k \lrcorner \dvol\,, 
\end{equation}
and, \nobreak
\begin{equation}
    \mathcal{E}_{\partial_u, \Sigma_{\tau}}(\phi) = \frac{1}{\sqrt{2}}\int_{\Sigma_{\tau}}\biggl[\vert \phi_{\hat{l}}\vert^2 + \frac{R^2F}{2}\vert \phi_{\hat{n}}\vert^2 + \left(1 + \frac{R^2F}{2}\right) \left( 2 Re\left(\nabla_{\hat{m}}\phi \nabla_{\hat{\Bar{m}}}\phi \right) + \frac{1}{12}\mathrm{Scal}_{\hat{g}}\phi^2 \right) \biggr] \deriv \Sigma_{\tau}\, . \label{energy_sigma_tau}
\end{equation}
Furthermore we have : 
\begin{equation}
    \nabla_{\hat{m}}\phi \nabla_{\hat{\bar{m}}}\phi = \frac{1}{2}\left(\vert\partial_{\theta}\phi\vert^2 + \frac{1}{\sin^{2}{\theta}}\vert\partial_{\varphi}\phi\vert^2 \right) = \vert \nabla_{S^2}\phi\vert^2\, , 
\end{equation}
hence $Re\left(\nabla_{\hat{m}}\phi \nabla_{\hat{\Bar{m}}}\phi \right)$ is clearly a non negative quantity. 
\subsection{Error terms and energy estimates on the Vaidya spacetime}
\label{ss_EE}
\subsubsection{On the geometrical framework}
Our approach to the scattering is based on energy estimates. Stokes’ Theorem will allow us to obtain equalities and inequalities between the energy fluxes through our different hypersurfaces, from a conservation law for the energy current $J_a = V^b T_{ab}$. 
There will be error terms coming from the non-zero divergence of the energy current :
\begin{equation*}
    \nabla^a J_a = \nabla^{(a} \left(V^{b)}T_{ab}\right) = \nabla^{(a} V^{b)} T_{ab} + V^b \nabla^a T_{ab} \, .
\end{equation*}
Finally we obtain (see Appendix \ref{app-vaidya_error} for more details), for any smooth mass function $m(u)$ :
\begin{equation*}
    \nabla^{(a}V^{b)} = -m'(u)R^3\hat{n}^{(a} \hat{n}^{b)}\, .
\end{equation*}
The divergence of the energy momentum tensor $T_{ab}$ given in \eqref{energy-tensor} is : 
\begin{equation*}
    \nabla^aT_{ab} = \square_{\hat{g}}\phi \nabla_b \phi + \nabla_b \left(\frac{1}{12} \mathrm{Scal}_{\hat{g}}\phi^2\right) = \frac{1}{12}\nabla_b \left(\mathrm{Scal}_{\hat{g}}\right) \phi^2\, .
\end{equation*}
Provided that $\phi$ satisfies \eqref{eq_ondes_rescaled} and using $\mathrm{Scal}_{\hat{g}} = 12m(u)R$, we have :
\begin{equation}
    \nabla^a J_a = m'(u) R^3 \vert \phi_{\hat{n}} \vert^2 + m'(u)R \phi^2\, . \label{div-J}
\end{equation}
Taking $\mathcal{B}$ a domain closed by two hypersurfaces : $\mathcal{S}_{t_2}$ and $\mathcal{S}_{t_1}$ and foliated by $\mathcal{S}_t$. The equation \eqref{div-J} and Stokes' Theorem give us the following identity for any scalar field $\phi$ on $\mathcal{B}$ :
\begin{equation} 
\mathcal{E}_{\mathcal{S}_{t_2}} (\phi)  - \mathcal{E}_{\mathcal{S}_{t_1}} (\phi)  = \int_{\mathcal{B}} \nabla^a J_a \mathrm{dVol}^4 \, .\label{EnId2}
\end{equation}
\noindent The right-hand side of \eqref{EnId2} can be decomposed into an integral in $\tau$ over $[\tau_1,\tau_2]$ of integrals over $\Sigma_{\tau}$.  This is done by splitting the $4$-volume measure using the identifying vector field $\nu$ as follows
\[ \mathrm{dVol}^4 = \deriv \tau \wedge \nu\lrcorner \mathrm{dVol}^4 \, .\]

\begin{figure}[!ht]
    \centering
    \includegraphics[width=10cm]{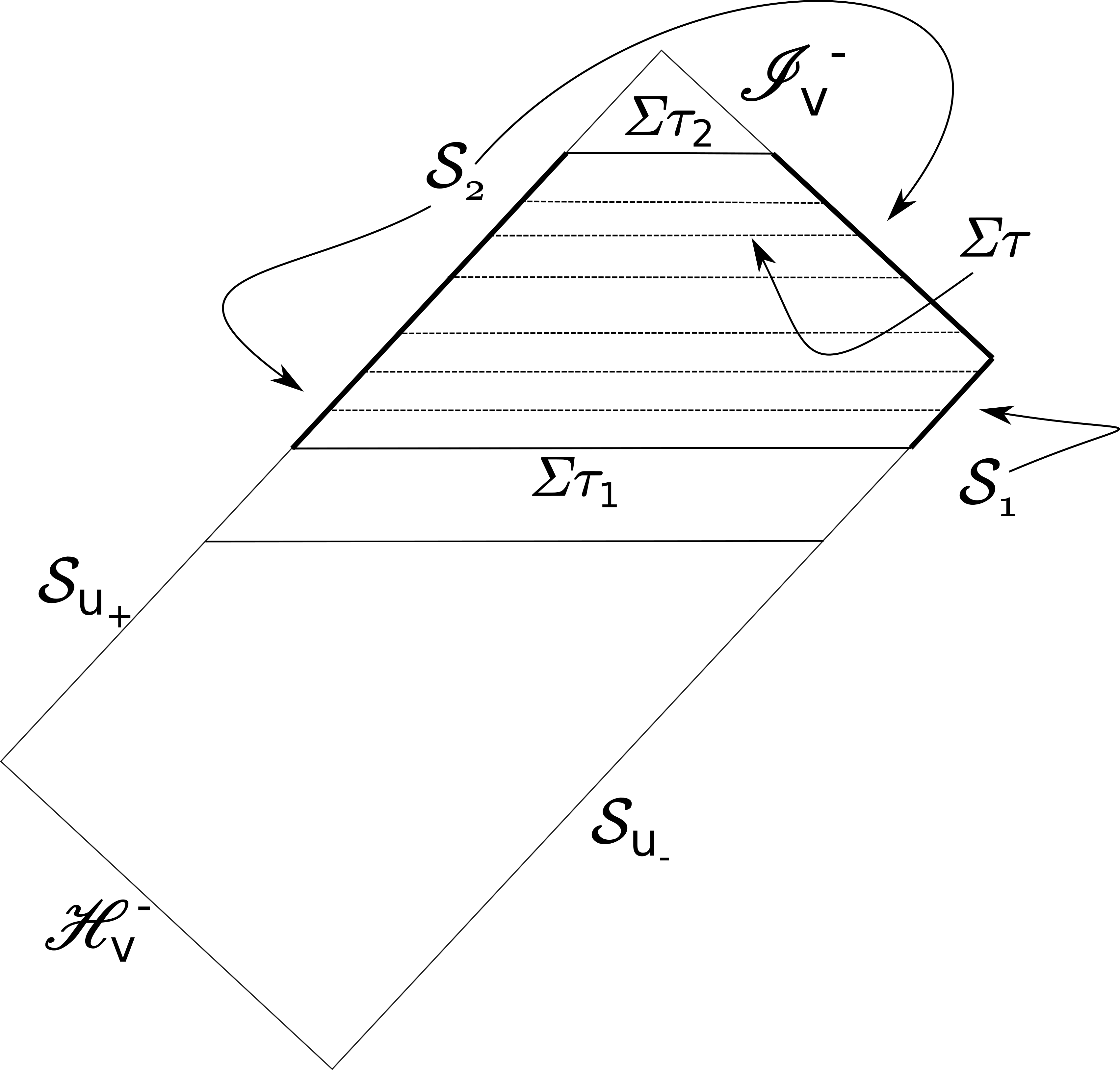}
    \caption{illustration of the foliation by $\Sigma_{\tau}$ of the Vaidya area.}
    \label{fig:V_feuilletage}
\end{figure}
We decompose the proof in two parts. Firstly we focus on the equivalence in the past \eqref{approx-vaidya-past} and secondly in the future \eqref{aprox-vaidya-future}. In the past, we consider $\tau_- \leq \tau_1 \leq \tau \leq \tau_2 \leq 0$ and the boundary is made of : $\Sigma_{\tau_1}, \Sigma_{\tau_2}, \mathcal{S}_1$ and $\mathcal{S}_2$ : 
\begin{align*}
    \mathcal{S}_1 =& \big(\mathscr{H}^{-}_V \cup \mathcal{S}_{u_-}\big)\cap \lbrace \tau_1 \leq \tau \leq \tau_2 \rbrace \, , \\
    \mathcal{S}_2=& \big(\mathcal{S}_{u_+}\cup \mathscr{I}^+_{\mathrm{V}}\big)\cap \lbrace \tau_1 \leq \tau \leq \tau_2 \rbrace \, . 
\end{align*}
Stoke's Theorem gives : 
\begin{equation}\label{Stockes_Vaidya}
    \mathcal{E}_{\Sigma_{\tau_2}}(\phi) + \mathcal{E}_{S_2}(\phi) - \mathcal{E}_{\Sigma_{\tau_1}}(\phi) - \mathcal{E}_{S_1}(\phi) = \int_{\tau_1}^{\tau_2} \int_{\Sigma_{\tau}} \mathrm{Err}(\phi) \deriv \Sigma \deriv \tau\, . 
\end{equation}
where
\begin{equation} \label{ErrTerm}
    \mathrm{Err}(\phi) =m'(u)R^3 \vert \phi_{\hat{n}}\vert^2 + m'(u)R\phi^2\, , 
\end{equation}
From this we infer : 
\begin{equation}
    \vert  \mathrm{Err}(\phi)\vert \leq \vert m'(u)\vert R^3 \vert \phi_{\hat{n}}\vert^2 + \vert m'(u) \vert R \phi^2\, .
\end{equation}
Remarking that the term $Re(\nabla_{\hat{m}} \phi \nabla_{\hat{\bar{m}}}\phi) = \vert\nabla_{S^2}\phi\vert^2\geq 0$ and that $m'(u)$ is a bounded function on $V$, there exists a positive constant $C$ such that : 
\begin{equation}
    \vert \mathrm{Err}(\phi) \vert \leq C \left[\vert \phi_{\hat{l}}\vert^2 + \frac{R^2F}{2}\vert \phi_{\hat{n}}\vert^2 + \left(1 + \frac{R^2F}{2}\right) \left( 2 Re\left(\nabla_{\hat{m}}\phi \nabla_{\hat{\Bar{m}}}\phi \right) + \frac{1}{12}\mathrm{Scal}_{\hat{g}}\phi^2 \right) \right]\, .
\end{equation}
The right hand side corresponds up to a constant to the integral of the energy flux across $\Sigma_{\tau}$ given in  \eqref{energy_sigma_tau}. As a direct consequence we have : 
\begin{equation*}
    \int_{\Sigma_{\tau}} \vert\mathrm{Err}(\phi)\vert \deriv u \wedge \deriv \omega \leq \cte \mathcal{E}_{\Sigma_{\tau}}(\phi)\, . 
\end{equation*}
Then, by bounding the error term in this way, equation \eqref{Stockes_Vaidya} entails that two estimates : 
\begin{align*}
    \mathcal{E}_{\Sigma_{\tau_2}}(\phi) + \mathcal{E}_{S_2}(\phi) - \mathcal{E}_{\Sigma_{\tau_1}}(\phi) - \mathcal{E}_{S_1}(\phi) \leq &\, \cte\int_{\tau_1}^{\tau_2} \mathcal{E}_{\Sigma_{\tau}}(\phi) \deriv \tau\, , \\
    \mathcal{E}_{\Sigma_{\tau_1}}(\phi) + \mathcal{E}_{S_1}(\phi) - \mathcal{E}_{\Sigma_{\tau_2}}(\phi) - \mathcal{E}_{S_2}(\phi) \leq &\, \cte\int_{\tau_1}^{\tau_2} \mathcal{E}_{\Sigma_{\tau}}(\phi) \deriv \tau\, . 
\end{align*}
Knowing that $S_1$ and $S_2$ are reunion of null hypersurfaces, the dominant energy condition entails that $\mathcal{E}_{S_1}(\phi)$ and $\mathcal{E}_{S_2}(\phi)$ are non negative. Then we obtain :
\begin{align}
    \mathcal{E}_{\Sigma_{\tau_2}}(\phi) + \mathcal{E}_{S_2}(\phi) - \mathcal{E}_{\Sigma_{\tau_1}}(\phi) - \mathcal{E}_{S_1}(\phi) \leq &\, \cte\int_{\tau_1}^{\tau_2} \left(\mathcal{E}_{\Sigma_{\tau}}(\phi) + \mathcal{E}_{S_2}(\phi)\right) \deriv \tau\, , \\
    \mathcal{E}_{\Sigma_{\tau_1}}(\phi) + \mathcal{E}_{S_1}(\phi) - \mathcal{E}_{\Sigma_{\tau_2}}(\phi) - \mathcal{E}_{S_2}(\phi) \leq &\, \cte\int_{\tau_1}^{\tau_2} \left(\mathcal{E}_{\Sigma_{\tau}}(\phi) + \mathcal{E}_{S_1}(\phi)\right) \deriv \tau\, . 
\end{align}
Using Grönwall's lemma on the bounded domain $[\tau_1, \tau_2]$, 
\begin{align}
     \mathcal{E}_{\Sigma_{\tau_2}}(\phi) + \mathcal{E}_{S_2}(\phi) \leq \cte \left( \mathcal{E}_{\Sigma_{\tau_1}}(\phi) + \mathcal{E}_{S_1}(\phi)\right) \, ,\label{estimate_vaidya_past1}\\ 
    \mathcal{E}_{\Sigma_{\tau_1}}(\phi) + \mathcal{E}_{S_1}(\phi) \leq \cte \left( \mathcal{E}_{\Sigma_{\tau_2}}(\phi) + \mathcal{E}_{S_2}(\phi)\right) \, . \label{estimate_vaidya_past2}
\end{align}
Taking $\tau_1 =\tau_-$ and $\tau_2 = 0$, hypersurfaces $\mathcal{S}_1$ and $\mathcal{S}_2$ become  :
\begin{align}
     \Sigma_{\tau_1} + \mathcal{S}_1 =& \mathscr{H}^{-}_V \cup \mathcal{S}_{u_-}\cap \lbrace 
     t \leq 0\rbrace = \mathcal{B}_-^V  \, , \\
   \Sigma_{\tau_2} + \mathcal{S}_2=& \Sigma_0^V + \mathcal{S}_{u_+}\cap \lbrace t\leq 0 \rbrace = \Sigma_{0,-}^V \, . 
\end{align}
hence, \eqref{estimate_vaidya_past1} and \eqref{estimate_vaidya_past2} turn into :
\begin{align*}
     \mathcal{E}_{\Sigma_{0,-}^V}(\phi) \leq \cte \left( \mathcal{E}_{\mathcal{B}_-^V}(\phi)\right) \, ,\\
     \mathcal{E}_{\mathcal{B}_-^V }(\phi) \leq \cte \left( \mathcal{E}_{\Sigma_{0,-}^V}(\phi) \right) \, . 
\end{align*}
This concludes the proof for \eqref{approx-vaidya-past}. \qed

We apply the same reasoning in the future. Let $0\leq \tau_1\leq \tau \leq \tau_2 \leq \tau_+$ and : 
\begin{align*}
    \mathcal{S}_1 =& \mathscr{I}^{+}_V \cup \mathcal{S}_{u_+}\cap \lbrace \tau_1 \leq \tau \leq \tau_2 \rbrace \, , \\
    \mathcal{S}_2=& \mathcal{S}_{u_-}\cap \lbrace \tau_1 \leq \tau \leq \tau_2 \rbrace \, . 
\end{align*}
Then Stoke's theorem and Grönwall's lemma still hold and now taking $\tau_1 = 0, \tau_2 = \tau_+$ we get : 
\begin{align*}
    \Sigma_{\tau_1} + \mathcal{S}_2 =& \Sigma_0^V + \mathcal{S}_{u_-}\cap\lbrace t\geq 0 \rbrace  = \Sigma_{0,+}^V\, , \\
    \Sigma_{\tau_2} + \mathcal{S}_1 =& \mathscr{I}^+_V + \mathcal{S}_{u_+}\cap \lbrace t\geq 0\rbrace = \mathcal{B}_+^V \, . 
\end{align*}
This leads to : 
\begin{align*}
     \mathcal{E}_{\Sigma_{0,+}^V}(\phi) \leq \cte \left( \mathcal{E}_{\mathcal{B}_+^V}(\phi)\right) \, ,\\
     \mathcal{E}_{\mathcal{B}_+^V }(\phi) \leq \cte \left( \mathcal{E}_{\Sigma_{0,+}^V}(\phi) \right) \, . 
\end{align*}
and this concludes the proof for \eqref{aprox-vaidya-future}. \qed

\section{Conformal scattering}
\label{conformal scattering}
From the energy estimates obtained above, we construct the conformal scattering operator in the following manner : \begin{enumerate}
    \item On specified hypersurfaces (typically $\Sigma_t, \mathscr{I}^{\pm}, \mathscr{H}^{\pm}$), we define energy spaces that are the completion of the space of smooth compactly supported functions on these hypersurfaces in the norm given by the energy fluxes computed in section \ref{energy estimates} for a solution of \eqref{eq_ondes_rescaled}. 
    \item We define the future trace operator that to smooth and compactly supported initial data associates the future scattering data, i.e. the restriction of the solution on the future boundary. Using energy estimates obtained in Theorem \ref{th_estimates} this operator is then extended as a bounded linear operator, one-to-one, with closed range between energy spaces in Proposition \ref{prop_trace}. 
    \item In order to prove that the trace operator is an isomorphism and knowing the previous properties, all we need to prove is that its range is dense. We solve this problem following Hormänder in \cite{1990_Hormander} and Nicolas in \cite{2006_Nicolas}, \cite{2016_Nicolas} by solving a Goursat problem for the scattering data on the null hypersurfaces $\mathscr{H}^+\cup \mathscr{I}^+$. This leads to Theorem \ref{thm_isomorphism}. 
    \item Finally, we apply the same procedure to define the past trace operator and then obtain the scattering operator in Theorem \ref{thm_scat_isomorphism}, which is constructed as an isomorphism between the energy space on the past boundary and the energy space on the future boundary.
\end{enumerate}

\subsection{Trace operator and energy spaces}
\begin{definition} \label{def_Ht}
    \emph{(Energy space of initial data).} Let $\mathcal{H}_0$ be the completion of $\mathcal{C}^{\infty}_0(\Sigma_t)\times \mathcal{C}^{\infty}_0(\Sigma_t)$ in the norm :
    \begin{equation*}
         \Vert  (\phi_0, \phi_1) \Vert^2_{\mathcal{H}_0} = \frac{1}{\sqrt{2}}\int_{\Sigma_0}\left[\vert \phi_{\hat{l}}\vert^2 + \frac{R^2F}{2}\vert \phi_{\hat{n}}\vert^2 + \left(1 + \frac{R^2F}{2}\right) \left( 2 Re\left(\nabla_{\hat{m}}\phi \nabla_{\hat{\Bar{m}}}\phi \right) + \frac{1}{12}\mathrm{Scal}_{\hat{g}}\phi^2 \right) \right] \deriv \Sigma\, ,
    \end{equation*}
    or, equivalently in coordinates $(u,R,\theta, \varphi)$ :
    \begin{equation*}
        \Vert  (\phi_0, \phi_1) \Vert^2_{\mathcal{H}_0} = \int_{\Sigma_0} \left[ \phi_u^2 + \dfrac{R^2F}{4\psi}(\psi + 1) \left(\dfrac{R^2F}{\psi} \phi_R^2 + \vert \nabla_{S^2}\phi\vert^2 + 2m(u)R \phi^2\right) \right] \deriv u \wedge \deriv\omega \, ,
    \end{equation*}
    with the mass function $m(u)$ as defined in \eqref{def_mass}. The function $\psi$ is the function that appears in the Vaidya metric and satisfies \eqref{EDOphi}. In region I, it is clear that $\psi = 1$.  In region II, we have $F=1-2m_{+}/r$ and $\psi$ is a constant (see Remark \ref{psi_cte}).
\end{definition}

\begin{remark}
    \label{psi_cte} From \cite{2021_Coudray-Nicolas}, we know that $m'(u)= 0$ ensures that $\psi$ is a constant function along incoming null principal geodesics. Thus on region II, we have : 
    \begin{equation*}
        \frac{\deriv }{\deriv u}( \psi \circ \gamma) = 0 \Rightarrow \psi = \cte\,,
    \end{equation*}
    where $\gamma = \gamma(u) = (u,r(u), \omega)$ is an incoming principal geodesic. One the other hand, it is clear that along outgoing null geodesics, $\psi$ is also a constant function, hence we conclude that on region II, $\psi = \cte$. 
\end{remark}

\begin{definition}\label{def-H+}
We define on $\mathscr{H}^+\cup \mathscr{I}^+$ the function space for scattering data $\mathcal{H}^+$ as the completion of $\mathcal{C}^{\infty}_{0}(\mathscr{H}^+) \times \mathcal{C}^{\infty}_{0}(\mathscr{I}^+)$ in the norm
\begin{equation*}
    \Vert (\xi, \zeta) \Vert^2_{\mathcal{H}^+ } = \frac{1}{2}\left(\int_{\mathscr{H}^+} \xi_v^2 \,\deriv v \wedge \deriv \omega +\int_{\mathscr{I}^+} \zeta_u^2 \,\deriv u \wedge \deriv \omega\right) \, . 
\end{equation*}
\end{definition}
\noindent We define the future trace operator for the solution $\phi$ of \eqref{eq_ondes_rescaled} with initial data $\big(\phi_0 = \phi\vert_{\Sigma_0}, \phi_1 = \partial_t\phi\vert_{\Sigma_0}\big)$, as the operator that associates to the initial data, the trace of the solution on the future-part of the boundary $\partial \hat{\mathcal{M}}$, i.e. $\left( \phi\vert_{\mathscr{H}^+}, \phi\vert_{\mathscr{I}^{+}}\right)$. 
This can be justified using Leray's theorem (see \cite{1953_Leray}) that ensures that the solution to \eqref{eq_ondes_rescaled} associated to initial data $(\phi_0, \phi_1)\in \mathcal{H}_0$ exists and is unique.  
\begin{definition} \emph{(Future trace operator).}
    Let $(\phi_0, \phi_1)\in \mathcal{C}^{\infty}_0(\Sigma_0)\times \mathcal{C}^{\infty}_0(\Sigma_0)$. Consider $\phi \in \mathcal{C}^{\infty}(\hat{\mathcal{M}})$ the solution of \eqref{eq_ondes_rescaled} such that :
    \begin{equation*}
        \phi\vert_{\Sigma_0} = \phi_0, \; \partial_t \phi\vert_{\Sigma_0} = \phi_1\, . 
    \end{equation*}
    We define the trace operator $\mathcal{T}^+$ from $\mathcal{C}^{\infty}_0(\Sigma_0)\times \mathcal{C}^{\infty}_0(\Sigma_0)$ to $\mathcal{C}^{\infty}(\mathscr{H}^+)\times \mathcal{C}^{\infty}(\mathscr{I}^+)$ as follows :
    \begin{equation*}
        \mathcal{T}^{+}(\phi_0, \phi_1) = \left(\phi\vert_{\mathscr{H}^+}, \phi\vert_{\mathscr{I}^{+}}\right)\, . 
    \end{equation*}
\end{definition}

From Theorem \ref{th_estimates}, we infer the following proposition (the same proposition holds for the past trace operator) :  
\begin{proposition}\label{prop_trace}
The trace operator $\mathcal{T}^{+}$ extends uniquely as a bounded linear map from $\mathcal{H}_t$ to $\mathcal{H}^{+}$ that still satisfies :
\begin{equation*}
    \Vert \mathcal{T}^{+}(\phi_0,\phi_1)\Vert_{\mathcal{H}^+} \simeq \Vert (\phi_0, \phi_1)\Vert_{\mathcal{H}_0} \, , 
\end{equation*}
 is one-to-one and that its range is closed. 
\end{proposition}
\textbf{Proof :} It is clear that $\mathcal{T}^+$ is a linear operator, since the propagation equation \eqref{eq_ondes_rescaled} is linear; it acts between Hilbert spaces $\mathcal{H}_0$ and $\mathcal{H}^+$. Furthermore,  
\begin{equation*}
    \Vert \mathcal{T}^{+}(\phi_0,\phi_1)\Vert_{\mathcal{H}^+} \simeq \Vert (\phi_0, \phi_1)\Vert_{\mathcal{H}_0} \,,  \text{ i.e. }  C_1 \Vert (\phi_0, \phi_1)\Vert_{\mathcal{H}_0} \leq \Vert \mathcal{T}^{+}(\phi_0,\phi_1)\Vert_{\mathcal{H}^+} \leq C_2 \Vert (\phi_0, \phi_1)\Vert_{\mathcal{H}_0}
\end{equation*}
with $C_1<C_2$ two real positive constants. 
It is now a well-known theorem (see Theorem 2.5 in \cite{2002_Abramovich}) that a such operator is one-to-one and has closed range. \qed

\subsection{Goursat problem and Scattering operator}
In order to prove that the future trace operator is an isomorphism we need to prove that the range of $\mathcal{T}^+$ is dense in $H^1(\mathscr{I}^+)$. We follow the method described by Nicolas for the Schwarzschild spacetime in \cite{2016_Nicolas} by solving the Goursat problem from $\mathscr{I}^+\cup \mathscr{H}^+$ for data in $\mathcal{C}^{\infty}_0(\mathscr{I}^+)\times \mathcal{C}^{\infty}_0(\mathscr{H}^+)$. This way, the support of data on $\mathscr{H}^+$ and $\mathscr{I}^+$ remains away from $i^+$. In this context we state the following proposition : 
\begin{proposition}\label{prop_goursat}
    Let $(\phi^{\infty}_{\mathscr{H}^{+}}, \phi^{\infty}_{\mathscr{I}^+}) \in \mathcal{C}^{\infty}_{0}(\mathscr{H}^+) \times \mathcal{C}^{\infty}_{0}(\mathscr{I}^+)$, there exist :  
    \begin{equation*}
        \left(\phi_{\Sigma_0}, \partial_t\phi_{\Sigma_0} \right) \in \mathcal{H}_0 \, ,
    \end{equation*}
    such that :
    \begin{equation*}
        (\phi^{\infty}_{\mathscr{H}^{+}}, \phi^{\infty}_{\mathscr{I}^+}) = \mathcal{T}^{+}\left(\phi_{\Sigma_0}, \partial_t\phi_{\Sigma_0} \right)\, ,
    \end{equation*}
    With $\mathcal{H}_0$ the energy space of initial data at $t=0$, (see definition \ref{def_Ht}). 
\end{proposition}
\textbf{Proof : } First and foremost, we remark that the singularity at $i^+$ is completely avoided here, then we will focus on the singularity $i^0$ on $\mathscr{I}^+$. The compact support of $\phi^{\infty}_{\mathscr{I}^+}$ on $\mathscr{I}^+$ ensures that Hormänder's results in \cite{1990_Hormander} hold. Note that Hormänder dealt with a spatially compact spacetime, that is not the case here due to the singularity of the conformal boundary at $i^0$. However, according to \cite{2016_Nicolas} in Appendix B, it is possible to extend Hormänder's results to the conformal compactified Schwarzschild spacetime. The construction proceeds as follows: data on $\mathscr{H}^+ \cup \mathscr{I}^+$ are compactly supported to ensure that their past support remains away from $i^+$. Let $\mathcal{S}$ be a spacelike hypersurface on $\bar{\mathcal{M}}$ that intersects $\mathscr{I}^+$ and $\mathscr{H}^+$ in the past of the supported data. Then we consider the future of this hypersurface, denoted by $\mathcal{J}^+(\mathcal{S})$, and we remove the future of a point lying in the future of the past support of the data. The resulting spacetime is finally extended as a cylindrical globally hyperbolic spacetime. Within this framework, Nicolas proved the uniqueness of the solution of the Goursat problem in the future of $\mathcal{S}$. With this construction, we can apply Hormänder's results in our study:
\begin{proposition}{(Hormänder, 1990)}

\noindent Let $\mathcal{S}$ be a spacelike hypersurface on the rescaled spacetime, that either crosses $\mathscr{I}^+$ in the past of the support of $\phi^{\infty}_{\mathscr{I}^+}$ and $\mathscr{H}^+$ in the past of the support of $\phi^{\infty}_{\mathscr{H}^+}$. We denote by $\mathcal{J}^+(\mathcal{S})$ the causal future of $\mathcal{S}$. Then there exists a unique solution $\phi$ of \eqref{eq_ondes_rescaled} such that : 
\begin{enumerate}
    \item $\phi \in H^1(\mathcal{J}^+(\mathcal{S}))$. 
    \item $\phi\vert_{\mathscr{I}^+} = \phi_{\mathscr{I}^+}^{\infty}$ and $\phi\vert_{\mathscr{H}^+} = \phi_{\mathscr{H}^+}^{\infty}$. 
\end{enumerate}
\end{proposition}
\begin{figure}[!ht]
    \centering
    \includegraphics[scale=0.5]{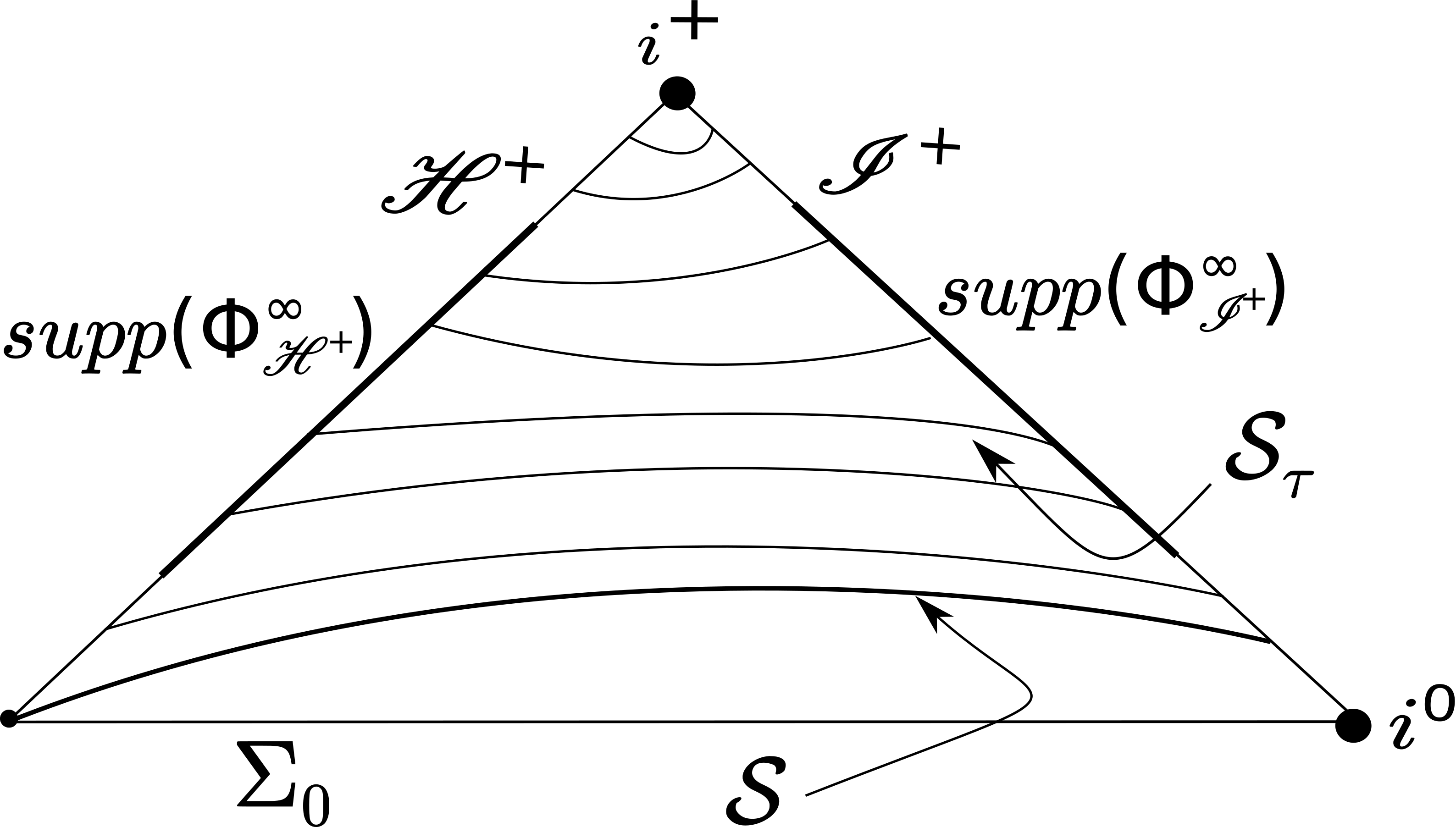}
    \caption{Foliation of the future of $\mathcal{S}$ by spacelike hypersurfaces $\mathcal{S}_{\tau}$ transverse to the future null boundary. }
\end{figure}
Now we can propagate the solution $\phi$ in the past, down to $\Sigma_0$ in a way that avoids the spacelike infinity $i^0$. 

Due to our construction, with $\mathcal{S}$ in the past of the support of $\phi^{\infty}_{\mathscr{H}^+}$ and $\phi^{\infty}_{\mathscr{I}^+}$, the solution $\phi\vert_{S}$ vanishes at the boundary, i.e. at $\mathcal{S}\cap \mathscr{I}^+$ and at $\mathcal{S}\cap\mathscr{H}^+$. Then $\big(\phi\vert_{\mathcal{S}}, \partial_t\phi\vert_{\mathcal{S}}\big)$ is naturally in $\big(H^1_0(\mathcal{S}), L^2(\mathcal{S})\big)$. We infer from this that we can define two sequences of smooth functions : $\big(\phi^n_{0,\mathcal{S}}\big)_n$ and $\big(\phi^n_{1,\mathcal{S}}\big)_n$ that converge in the following manner : 
\begin{align*}
    \phi^n_{0,\mathcal{S}} &\overset{H^1_0(\mathcal{S})}{\longrightarrow} \phi\vert_{\mathcal{S}}\, , \\
     \phi^n_{1,\mathcal{S}} &\overset{L^2(\mathcal{S})}{\longrightarrow} \partial_t\phi\vert_{\mathcal{S}}\, .
\end{align*}
Consider now $(\phi^n)_n$ the sequence of smooth solutions of \eqref{eq_ondes_rescaled} on the rescaled spacetime with initial data $\left(\big(\phi^n_{0,\mathcal{S}}\big)_n,\big(\phi^n_{1,\mathcal{S}}\big)_n\right)$ on $\mathcal{S}$. Because firstly the support of $\phi^n$ is compact on $\mathcal{S}$ and secondly because the conformal metric remains bounded on $\mathrm{supp}(\phi^n\vert_{\mathcal{S}})$, then all the weights that appears in the energy flux $\mathcal{E}_{\partial_t, \mathcal{S}}(\phi^n)$ are bounded and : 
\begin{equation}
    \mathcal{E}_{\partial_t, \mathcal{S}}(\phi^n) \simeq \Vert\phi^n\Vert^2_{H^1_0(\mathcal{S})}\, . 
\end{equation}
Furthermore, due to the compact support of the solution on $\mathcal{S}$ and knowing the finite speed propagation of scalar waves, this ensures that $\phi$ vanishes in a neighbourhood of $i^0$. Thus, using Theorem \ref{th_estimates}, we obtain the following equivalence between energy fluxes associated to the observer $\partial_t$ on $\mathcal{S}$ and $\Sigma_0$ : 
\begin{equation}
    \mathcal{E}_{\partial_t, \mathcal{S}}\big(\phi^n\big) \simeq \mathcal{E}_{\partial_t, \Sigma_0}\big(\phi^n\big)\, .  \label{estimates_phi_n}
\end{equation}
Denoting : 
\begin{equation*}
    \phi_{\Sigma_0} = \phi \vert_{\Sigma_0}\, \text{ and }\, \partial_t\phi_{\Sigma_0} = \partial_t\phi\vert_{\Sigma_0}\, , 
\end{equation*}
the equivalence \eqref{estimates_phi_n} leads to : 
\begin{equation*}
    \big(\phi^n_{\Sigma_0},\partial_t\phi^n_{\Sigma_0}\big) \overset{\mathcal{H}_0}{\longrightarrow} \big(\phi\vert_{\Sigma_0}, \partial_t\phi\vert_{\Sigma_0}\big)\, .
\end{equation*}
Hence : 
\begin{equation*}
    \left(\phi_{\Sigma_0}, \partial_t\phi_{\Sigma_0}\right) \in \mathcal{H}_0\, , 
\end{equation*}
and this satisfies : 
\begin{equation*}
    \left(\phi^{\infty}_{\mathscr{H}^+}, \phi^{\infty}_{\mathscr{I}^+}\right) = \mathcal{T}^{+}\left(\phi_{\Sigma_0}, \partial_t\phi_{\Sigma_0}\right)\, . \qed
\end{equation*}
From Proposition \ref{prop_goursat} we infer that the range of $\mathcal{T}^{+}$ is dense in $\mathcal{H}^+$. Adding to this the Proposition \ref{prop_trace} we obtain the following theorem :
\begin{theorem}\label{thm_isomorphism}
    The trace operator $\mathcal{T}^+$ :
\begin{equation*}
    \begin{array}{rrcl}
        \mathcal{T}^+:&\mathcal{H}_0&\longrightarrow& \mathcal{H}^+\\
        &(\phi_0,\phi_1) &\longmapsto &(\phi\vert_{\mathscr{H}^+}, \phi\vert_{\mathscr{I}^+})\,,
    \end{array}
\end{equation*}
is an isomorphism. 
\end{theorem}

\begin{remark} 
    The same result holds for $\mathcal{T}^-$. Let $\phi$ be a solution of \eqref{eq_ondes_rescaled} with initial data $\big(\phi_0 = \phi\vert_{\Sigma_0}, \phi_1 = \partial_t\phi\vert_{\Sigma_0}\big)$, then the pas trace operator is defined as: 
    \begin{equation*}
        \begin{array}{rrcl}
        \mathcal{T}^-:&\mathcal{C}^{\infty}_0(\Sigma_0)\times \mathcal{C}^{\infty}_0(\Sigma_0)&\longrightarrow& \mathcal{C}^{\infty}(\mathscr{H^-})\times \mathcal{C}^{\infty}(\mathscr{I}^-)\\
        &(\phi_0,\phi_1) &\longmapsto &(\phi\vert_{\mathscr{H}^-}, \phi\vert_{\mathscr{I}^-})\,.
    \end{array}
    \end{equation*}
    It follows from the second approximate conservation law in Theorem \ref{th_estimates} that $\mathcal{T}^-$ extends as a bounded linear map : 
     \begin{equation*}
        \begin{array}{rrcl}
        \mathcal{T}^-:&\mathcal{H}_0&\longrightarrow& \mathcal{H}^-\\
        &(\phi_0,\phi_1) &\longmapsto &(\phi\vert_{\mathscr{H}^-}, \phi\vert_{\mathscr{I}^-})\,,
    \end{array}
    \end{equation*}
    satisfies : 
    \begin{equation*}
        \Vert \mathcal{T}^-(\phi_0, \phi_1)\Vert_{\mathcal{H}^-}\simeq \Vert (\phi_0, \phi_1)\Vert_{\mathcal{H}_0}\, ,
    \end{equation*}
    hence is one-to-one and has closed range. The resolution of Goursat problem for data compactly supported on $\mathscr{H}^- \cup \mathscr{I}^-$ is similar to what was done for the resolution of the Goursat problem on the future null boundary and entails that $\mathcal{T}^-$ is a isomorphism. 
\end{remark}

\begin{theorem}{\emph{(Scattering operator)} :}\label{thm_scat_isomorphism}

\noindent Let $\phi$ be the solution of \eqref{eq_ondes_rescaled} with initial data $(\phi_0,\phi_1) \in \mathcal{H}_0$. Consider $\left(\phi^{\infty}_{\mathscr{H}^{\pm}}, \phi^{\infty}_{\mathscr{I}^{\pm}}\right) \in \mathcal{H}^{\pm}$ the trace of $\phi$ on the respectively future and past boundary of the rescaled spacetime. Then the scattering operator $S$ obtained from future and past trace operators : 
    \begin{equation*}
        S = \mathcal{T}^+ \big(\mathcal{T}^-\big)^{-1}\, ,
    \end{equation*}
    that acts as : 
    \begin{equation*}
         \begin{array}{rrcl}
        S:&\mathcal{H}^-&\longrightarrow& \mathcal{H}^+\\
        &\big(\phi^{\infty}_{\mathscr{H}^-},\phi^{\infty}_{\mathscr{I}^-}\big) &\longmapsto &(\phi^{\infty}_{\mathscr{H}^+}, \phi^{\infty}_{\mathscr{I}^+})\,,
    \end{array}
    \end{equation*}
    is an isomorphism. 

\end{theorem}

\appendix
\section{Error terms in the Vaidya spacetime}\label{app-vaidya_error}
\subsection{Definition of the spin coefficients}
In this article we work with a Newman-Penrose tetrad $(\hat{l}, \hat{n}, \hat{m}, \hat{\bar{m}})$ that satisfies : 
\begin{equation*}
    \hat{l}^a \hat{n}_a = -\hat{m}^a\hat{\bar{m}}_ a = 1\,. 
\end{equation*}
Because of this normalisation there are only 12 independent spin coefficients to compute :
\begin{align*}
    \kappa =& m^a Dl_a\,, & \varepsilon =& \frac{1}{2}\left( n^a Dl_a + m^a D \bar{m}_a\right)\,, & \pi =& -\bar{m}^a Dn_a\,.\\
    \rho =& m^a \delta' l_a\,, & \alpha =& \frac{1}{2}\left( n^a \delta'l_a + m^a \delta' \bar{m}_a\right)\,, & \lambda =& -\bar{m}^a \delta'n_a\, .\\
    \sigma =& m^a \delta l_a\,, & \beta =& \frac{1}{2}\left( n^a \delta l_a + m^a \delta \bar{m}_a\right)\,, & \mu =& -\bar{m}^a \delta n_a\, .\\
    \tau =& m^a \Delta l_a\,, & \gamma =& \frac{1}{2}\left( n^a \Delta l_a + m^a \Delta \bar{m}_a\right)\,, & \nu =& -\bar{m}^a \Delta n_a\, .\\
\end{align*}
The other spin coefficients are $\alpha', \beta', \gamma'$ and $\varepsilon'$ and they are related to the previous coefficients with : 
\begin{equation*}
    \alpha' = -\beta\, , \beta' = - \alpha\,, \gamma' = -\varepsilon\, , \varepsilon'=-\gamma\, . 
\end{equation*}
\subsection{Proof in the Vaidya spacetime}
Here are the details to obtain \eqref{div-J}. We recall that we use the following tetrad : 
\begin{align}
    \hat{l}^a =& \partial_u^a + \frac{R^2F}{2}\partial_R^a\, ,\\
    \hat{n}^a =& - \partial_R^a\, ,\\
    \hat{m}^a =& \frac{1}{\sqrt{2}}\left(\partial_{\theta}^a + \frac{i}{\sin{\theta}}\partial_{\varphi}^a\right)\, , \hat{\bar{m}}^a =\frac{1}{\sqrt{2}}\left(\partial_{\theta}^a - \frac{i}{\sin{\theta}}\partial_{\varphi}^a\right)\,\, ,
\end{align}
The only non-zero spin coefficients are : 
\begin{equation*}
    \varepsilon = \frac{1}{2}R - \frac{3}{2}R^2m(u)\, , \alpha = -\beta = -\frac{\sqrt{2}}{4}\cot{\theta}\, . 
\end{equation*}
The Killing form of $T = \partial_u$ reads :
\begin{equation}
    \nabla^{(a}\partial_u^{b)} = \nabla^{(a} \hat{l}^{b)} + \nabla^{(a}\left(\Omega^2\frac{F}{2}\hat{n}^{b)}\right)\, . \label{killing_form}
\end{equation}
We decompose the connection along null directions of the tetrad : 
\begin{equation}
    \nabla^a = \hat{l}^a \Delta + \hat{n}^a D - \hat{m}^a\delta' - \hat{\Bar{m}}^a \delta\, . 
\end{equation}
The spin coefficient-equations for $\hat{l}$ and $\hat{n}$ are given by (see section 4.5 in \cite{1984_Penrose-Rindler}) :
\begin{align*}
    D\hat{l}^a = & (\varepsilon + \bar{\varepsilon})\hat{l}^a - \bar{\kappa}\hat{m}^a - \kappa \hat{\bar{m}}^a\, , & D\hat{n}^a =& (\gamma' + \bar{\gamma}')\hat{n}^a - \tau'\hat{m}^a - \bar{\tau}'\hat{\bar{m}}^a\, .  \\
    \delta\hat{l}^a = & (\beta + \bar{\alpha})\hat{l}^a - \bar{\rho}\hat{m}^a - \sigma \hat{\bar{m}}^a\, , & \delta\hat{n}^a =& (\alpha' + \bar{\beta}')\hat{n}^a - \rho'\hat{m}^a - \bar{\sigma}'\hat{\bar{m}}^a\, .  \\
    \delta'\hat{l}^a = & (\alpha + \bar{\beta})\hat{l}^a - \bar{\sigma}\hat{m}^a - \rho \hat{\bar{m}}^a\, , & \delta'\hat{n}^a =& (\beta' + \bar{\alpha}')\hat{n}^a - \sigma'\hat{m}^a - \bar{\rho}'\hat{\bar{m}}^a\, .  \\
    \Delta\hat{l}^a = & (\gamma + \bar{\gamma})\hat{l}^a - \bar{\tau}\hat{m}^a - \tau \hat{\bar{m}}^a\, , & \Delta\hat{n}^a =& (\varepsilon' + \bar{\varepsilon}')\hat{n}^a - \kappa'\hat{m}^a - \bar{\kappa}'\hat{\bar{m}}^a\, .  \\
\end{align*}
The only non-zero prime coefficients are $\alpha' = - \beta, \beta' = -\alpha$ and $\gamma' = - \varepsilon$. Furthermore, these coefficients are real. Thus the spin-coefficient equations with non-zero right hand side are :
\begin{align}
    D\hat{l}^a =& \, 2\varepsilon \hat{l}^a \, , \\
    D\hat{n}^a =& -2\varepsilon \hat{n}^a\, .
\end{align}
Equation \eqref{killing_form} becomes : 
\begin{equation}
    \nabla^{(a}\partial_u^{b)} = 2\varepsilon \hat{n}^{(a}\hat{l}^{b)} - \Omega^2 F\varepsilon \hat{n}^{(a}\hat{n}^{b)} + \nabla^{(a}\left(\frac{\Omega^2 F}{2}\right)\hat{n}^{b)}\, ,
\end{equation}
with $\Omega = R$ and: 
\begin{align*}
    \nabla^a F =& \hat{n}^a DF  + \hat{l}^a \Delta F = -\left(2m'(u) R - m(u)R^2F\right)\hat{n}^a + 2m(u) \hat{l}^a\, ,\\
    \nabla^a \Omega =&  \hat{n}^a D\Omega  + \hat{l}^a \Delta \Omega =  \frac{R^2F}{2}\hat{n}^a -\hat{l}^a\, .
\end{align*}
Finally we get : 
\begin{equation}
    \nabla^{(a}\partial_u^{b)} = -m'(u)R^3 \hat{n}^a \hat{n}^b\, .  
\end{equation}
\qed

\printbibliography
\end{document}